\documentclass[prx,twocolumn,superscriptaddress,floatfix,showpacs]{revtex4}
\usepackage{graphicx,amsfonts,amssymb,amsmath,hyperref,enumerate}

\newif\ifhyper
\hypertrue
\ifhyper
\hypersetup{
   citecolor = {green},
   colorlinks = {true}, 
   urlcolor = {blue} 
}
\fi

\newcommand{\beq}{\begin{equation}}
\newcommand{\eeq}{\end{equation}}
\newcommand{\beqa}{\begin{eqnarray}}
\newcommand{\eeqa}{\end{eqnarray}}
\newcommand{\ket} [1] {\vert #1 \rangle}
\newcommand{\bra} [1] {\langle #1 \vert}

\newcommand{\tr}{\mathop{\mathrm{tr}}}

\def\bra#1{\langle#1\vert}
\def\ket#1{\vert#1\rangle}

\def\Longarrow{\protect\@lra}
\def\@lra{\relbar\joinrel\relbar\joinrel\relbar\joinrel%
          \relbar\joinrel\rightarrow}

\usepackage{tikz}
\newcommand{\whitesite}{\begin{tikzpicture} \draw[black,fill=white] (0,0) circle (0.05); \end{tikzpicture}}
\newcommand{\blacksite}{\begin{tikzpicture} \draw[black,fill=black] (0,0) circle (0.05); \end{tikzpicture}}

\begin{document}

\title{Kitaev honeycomb tensor networks:Ê exact unitary circuits and applications}

\author{Philipp Schmoll}
\affiliation{Institute of Physics, Johannes Gutenberg University, 55099 Mainz, Germany}

\author{Rom\'an Or\'us}
\affiliation{Institute of Physics, Johannes Gutenberg University, 55099 Mainz, Germany}

\begin{abstract}

The Kitaev honeycomb model is a paradigm of exactly-solvable models, showing non-trivial physical properties such as topological quantum order, abelian and non-abelian anyons, and chirality. Its solution is one of the most beautiful examples of the interplay of different mathematical techniques in condensed matter physics. In this paper, we show how to derive a tensor network (TN) description of the eigenstates of this spin-1/2 model in the thermodynamic limit, and in particular for its ground state.  In our setting, eigenstates are naturally encoded by an exact 3d TN structure made of fermionic unitary operators, corresponding to the unitary quantum circuit building up the many-body quantum state. In our derivation we review how the different ``solution ingredients" of the Kitaev honeycomb model can be accounted for in the TN language, namely: Jordan-Wigner transformation, braidings of Majorana modes, fermionic Fourier transformation, and Bogoliubov transformation. The TN built in this way allows for a clear understanding of several properties of the model. In particular, we show how the fidelity diagram is straightforward both at zero temperature and at finite temperature in the vortex-free sector. We also show how the properties of two-point correlation functions follow easily. Finally, we also discuss the pros and cons of contracting of our 3d TN down to a 2d Projected Entangled Pair State (PEPS) with finite bond dimension. The results in this paper can be  extended to generalizations of the Kitaev model, e.g., to other lattices, spins, and dimensions. 

\end{abstract}


\maketitle

\section{Introduction}
\label{sec1}

In the study of strongly correlated systems and complex materials, one usually boils everything down to minimal, simple models,  in order to understand basic phenomena. Such models are greatly simplified but, still, most of the time they are just too hard to understand. This is the reason why we need clever numerical simulation methods, of which there are plenty of well-known examples. But sometimes we are lucky, and we can understand properties of some of these models \emph{exactly}. These are the so-called \emph{exactly-solvable models}, which have a longstanding tradition in statistical mechanics. Just to name some examples, Onsager's solution of the 2d classical Ising model \cite{onsager} was a monumental step forward in theoretical physics and  mathematics. Many important developments leading to exact solutions of classical statistical models were also due to Baxter \cite{Baxterbook}. From the quantum-mechanical perspective these results were also useful, given the correspondence between classical models in d spatial dimensions, and quantum models in (d-1) spatial dimensions \cite{Sachdev}. Another milestone development in the field of exactly solvable models was the Bethe ansatz \cite{Bethe}, and its connection to the Yang-Baxter equations \cite{Yangbaxter}. Many important theorems were also due to  Lieb, Schultz and Mattis \cite{LSM}. Interesting families of exactly-solvable models have been analyzed also by other techniques in the context of quantum many-body systems. For instance, the exact solution of the XY \cite{xy} and AKLT \cite{aklt} quantum spin chains (and their generalizations \cite{gen}) allowed a more precise understanding of quantum entanglement in condensed matter systems. Another example is the family of string-net models \cite{stringnet}, which account for all possible non-chiral topological phases of matter, and which correspond to renormalization group fixed points describing lattice gauge theories. Free boson and fermion lattice models are also exactly solvable, which has underpinned recent research in, e.g., topological insulators \cite{ti}. Finally, in the realm of quantum field theory, conformal field theories in 1+1 dimensions are also exactly solvable \cite{cft}, which turned out to be particularly useful also in the study of quantum many-body entanglement \cite{cftEnt}. 

In this context, Kitaev proposed another type of exactly solvable model in 2d which is nowadays commonly known as the \emph{Kitaev honeycomb model} \cite{KitaHon}. This is a model of spins 1/2 sitting on the sites of a honeycomb lattice, and interacting via highly-anisotropic nearest-neighbor interactions. The model can, in fact, be solved exactly in several ways \cite{KitaHon, Sol, Sol2}. These exact solutions show that there is an abelian phase, called A-phase, which hosts abelian anyons and $\mathbb{Z}_2$ topological order, being in the same universality class as the 2d Toric Code \cite{toriccode}. There is also a quantum phase transition to the so-called B-phase, which is gapless and with Dirac cones in the dispersion relation. The B-phase acquires a gap if one adds a small perturbation, like a magnetic field or a 3-spin interaction. In the presence of such a perturbation the phase is chiral and quasiparticle excitations are non-abelian Ising anyons. These results make this model a paradigmatic example to study abelian, non-abelian, chiral, and non-chiral topological phases, as well as quantum phase transitions between them, in a fully analytic way. Moreover, from the experimental point of view the model is relevant in the study of some materials, where one needs to consider also the presence of  competing Heisenberg and/or Hubbard interactions \cite{KitaHeis, KitaHubbard}. Such Hamiltonians are dubbed ``Kitaev-Heisenberg" and ``Kitaev-Hubbard" models, and since they are no longer exactly solvable, they have been the subject of several numerical studies in recent years. The Kitaev honeycomb model has also been generalized to different settings while keeping (most of) its exact solvability, e.g., to other 2d lattices using Clifford matrices instead of spin operators \cite{Clifford}, 3d lattices \cite{hermans}, and other values of the local spins \cite{otherspin}. Additionally, and from the perspective of quantum information theory, it was recently proven that the ground state of the model always obeys the so-called \emph{area-law} for the entanglement entropy of a block \cite{arealaw, arealawkita}. 

Independently of the above developments on exactly solvable models, Tensor Networks (TN) \cite{TN} have emerged recently as the natural language based on entanglement to understand quantum many-body states of matter. Many successful numerical algorithms have been developed using TNs, e.g., Density Matrix Renormalization Group (DMRG) \cite{dmrg} and Time-Evolving Block Decimation (TEBD) \cite{TEBD} in 1d, Projected Entangled Pair States (PEPS) in 2d \cite{PEPS}, Multi-Scale Entanglement Renormalization Ansatz (MERA) for critical systems \cite{MERA}, and many, many more \cite{TN}. But independently of numerics, TNs are also a very valuable tool from the analytical perspective, since they offer a unique visual picture of how quantum correlations build up quantum states of matter. In fact, exactly-solvable models usually have some exact TN structure behind, which accounts for the organization of correlations in the system. This, in turn, often provides useful insights into other non-trivial properties. There are, in fact, many well-known examples of this. For instance, string-net models have an exact TN representation of all their eigenstates in terms of PEPS and MERA \cite{TNstringnet}, which includes the Toric Code and the quantum double models as particular cases \cite{TCTN}. The Bethe anstaz also has a TN structure \cite{BetheTN}. Quadratic fermionic homogeneous Hamiltonians can be diagonalized via the fast Fourier transformation, which can be represented by the so-called spectral TN \cite{specTN}. Also, the ground state of the XY model can be written in terms of a 2d TN, which can be approximated by an MPS with exponentially good accuracy in the gapped phases \cite{XYTN}. 

On top of all this, during the last 3 years there has been a \emph{boom} in the study of chiral topological phases with TNs. While non-chiral topological phases in 2d admit a natural TN description (e.g. the string-net models mentioned above), the situation was not so clear for chiral topological phases, where chiral modes appear at the boundaries of the state. And this is relevant, since these phases are the ones entering in the description of very important strongly-correlated systems, such as fractional quantum Hall states \cite{fqhe} and topological insulators \cite{ti}. In this context, it was recently shown that one can build toy-model TN wavefunctions with chiral topological order using 2d PEPS, with the caveat that they also have infinite correlation length and are therefore the ground states of either (i) a gapless local Hamiltonian, or (ii) a gapped Hamiltonian with long-range interactions. Some of these wavefunctions were explicitly constructed for fermions, with parent Hamiltonians having fermi points in momentum space and a quadratic dispersion relation for quasiparticles \cite{Thorsten}.  

Given all the above, we have then the following natural questions: 

\begin{enumerate}
\item{What is the TN structure of the eigenstates of the Kitaev model, and what can be understood from it?}
\item{Can this TN be written, perhaps approximately, as a 2d PEPS with finite bond dimension?}
\end{enumerate}

Concerning the first question, we know that such a structure must exist, though it is not unique because different ways of solving the model will produce different TN structures for the eigenstates. Nevertheless, no matter which way of solving the model one chooses, finding such TN structure would help in better understanding the properties of the model and its generalizations. As for the second question, we expect this to be true at least close to the Toric Code regime (in the thermodynamic limit), where we know there is an exact PEPS representation. But the fact that the area-law for the entanglement entropy is always satisfied is giving us a hint that, perhaps, this may be possible in the whole parameter range of the model. In the B-phase, the PEPS constructed in this way would be a practical (and perhaps approximate) first example of a 2d PEPS with both infinite correlation length \emph{and} Dirac cones in the dispersion relation of quasiparticles that becomes chiral in the presence of a small perturbation. If however such a construction is not possible, then this would be the  first practical example of an area-law state that cannot be approximated by a 2d PEPS with finite bond dimension, which even if quite uncommon, it may also exist in principle \cite{Jens}. 

In this paper we fully answer the first question above, and partially answer the second. Specifically, we first provide an explicit and exact unitary 3d TN construction for the eigenstates of the Kitaev honeycomb model in the thermodynamic limit, and show how some well-known properties of the model can be derived straightforwardly from it, such as fidelity diagrams at zero and finite temperature. Then, we discuss the pros and cons of contracting such a 3d TN down to a 2d PEPS with finite bond dimension \footnote{As we shall explain, numerical simulations for systems up to 32 spins did not clarify the answer to question 2.}. In order to find the exact 3d TN structure we follow step-by-step a possible way of solving the Kitaev honeycomb model, and understand at each step how everything can be translated into the language of TNs. This may sound easy at first sight, but there are in fact derivations that turn out to be quite non-trivial, and which will for sure be useful for further applications of TNs in other contexts. Here we will focus, in particular, in the ground state of the system, but our 3d construction can be easily generalized to arbitrary eigenstates, as we shall also explain. Some of the properties of this 3d unitary TN will also be discussed, such as the existence of a causal cone. Moreover, we shall also prove that the ground state of the (fermionized) Kitaev Honeycomb model cannot be represented \emph{exactly} by a \emph{gaussian} fermionic 2d PEPS with finite bond-dimension, leaving the door open for other types of (perhaps approximate) PEPS. To the best of our knowledge, this work provides also the first explicit TN construction for the eigenstates of this paradigmatic exactly-solvable model of quantum matter. 

This paper is organized as follows. In Sec.\ref{sec2} we review some background material, namely: (i) Kitaev's honeycomb model, an exact solution using a Jordan-Wigner transformation, and the properties of its phase diagram; (ii) brief basics on TNs. In Sec.\ref{sec3} we construct the 3d unitary TN for the ground state of the model, and explain how this can be generalized to other eigenstates. We also explain here some of the basic properties of this TN construction, showing that it can be understood as a quantum circuit building up the quantum state (and hence has causal cones). In Sec.\ref{sec4} Êwe show our applications of this 3d TN to derive (i) the ground-state fidelity diagram, (ii) the thermal fidelity diagram in the vortex-free sector, and (iii) properties of the two-point correlation functions. In Sec.\ref{sec5} we discuss the possibility of approximating our 3d TN by a 2d PEPS with finite bond dimension in the whole parameter range, together with its potential implications. Finally, in Sec.\ref{sec6} we wrap up our conclusions, and provide some intuitions for further work along this direction. In Appendix \ref{appA} we explain a couple of technicalities concerning the calculation of the finite-size Hamiltonian that we use in our numerical checks, specially concerning the Fourier transformation and the boundary terms coming from the Jordan-Wigner transformation. 

\section{Background material}
\label{sec2}

\subsection{The Kitaev honeycomb model} 

\subsubsection{Hamiltonian}

\begin{figure}
\includegraphics[width=0.38\textwidth]{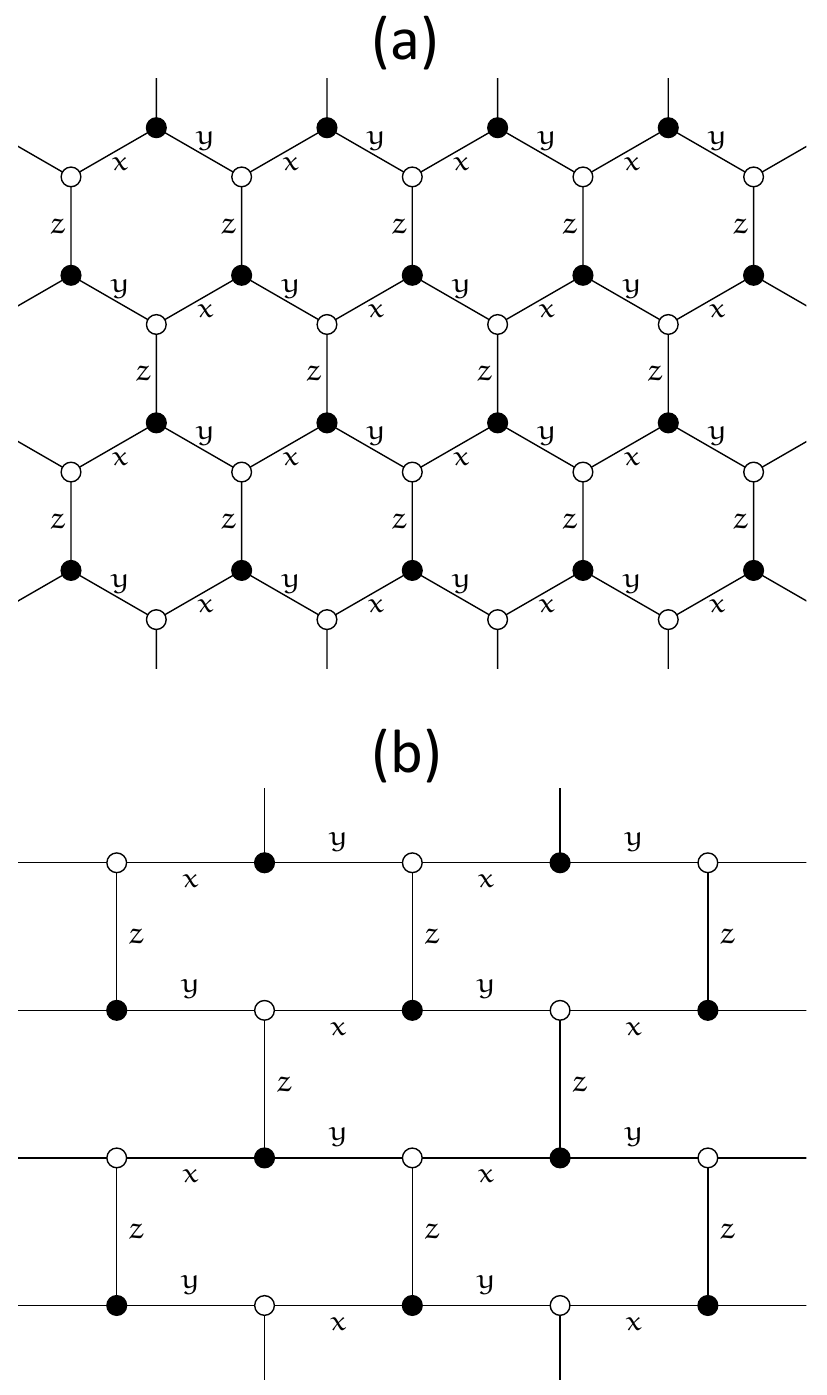}
\caption{(a) Original honeycomb lattice, and (b) topologically-equivalent brick wall lattice. The nearest-neighbor spin-spin interaction depends on the orientation of the bonds.}
\label{fig:KitaevHoneycombModel}
\end{figure}

The Kitaev honeycomb model, originally introduced by Kitaev in 2006 \cite{KitaHon}, describes a spin system with spins-1/2 at the sites of a two-dimensional honeycomb lattice. This lattice has connectivity three, which means that each site is connected to three nearest-neighbors. Additionally, the lattice is bi-colorable, see Fig.\ref{fig:KitaevHoneycombModel}, since it can be understood in terms of two overlapping triangular sublattices. The lattice, made of hexagons, is also topologically equivalent to a brick wall lattice. Some of the derivations in this paper are easier to visualize in this brick wall picture, so from now on we will stick to it. 

The model is defined by a nearest-neighbor interaction between the spins that depends on the bond orientation. More specifically, for the three different types of bonds $x$, $y$ and $z$ (see Fig.\ref{fig:KitaevHoneycombModel}) one has respectively $XX$, $YY$ and $ZZ$ Pauli interactions. The model is thus described by the Hamiltonian
\beq
	\hat{H} = -J_x \sum\limits_{x\text{-links}}\hat{\sigma}_i^{x}\hat{\sigma}_j^{x} 
		-J_y \sum\limits_{y\text{-links}}\hat{\sigma}_i^{y}\hat{\sigma}_j^{y}
		-J_z \sum\limits_{z\text{-links}}\hat{\sigma}_i^{z}\hat{\sigma}_j^{z} \ ,
		\label{kitaham} 
\eeq
with $\hat{\sigma}_i^\alpha$ the $\alpha$-Pauli matrix at site $i$. The physics of this Hamiltonian is symmetric with respect to permutations of the coupling strengths $J_x$, $J_y$ and $J_z$. As we shall review later, this Hamiltonian offers a variety of interesting phenomena, including different topological phases supporting both abelian and non-abelian anyons. The fact that it can host non-abelian anyons has made this model particularly interesting for topological quantum computing \cite{QcompAnyons}

\subsubsection{Exact solution}

The model can be solved exactly using different techniques. For instance, in the original proposal by Kitaev the model was mapped to a fermionic Hamiltonian with redundant degrees of freedom that were subsequently projected out \cite{KitaHon}. It is however possible to get around this projection by using a Jordan-Wigner transformation \cite{Sol2}. Another possibility involves a mapping to hard-core bosons \cite{Sol}. For the work in this paper we have chosen the second of these options, namely, the exact solution through a Jordan-Wigner transformation. We will see that this solution can be very neatly understood with the TN language. But before we do this, let us first review the formal solution of the model. 

\bigskip 
\underline{\emph{(i) Conserved quantities}}

An important feature of the Kitaev honeycomb model is the presence of infinitely-many conserved quantities, which is related to the fact that the model can be fully diagonalized exactly. Those conserved quantities are associated with the plaquettes $p$ of the lattice. For every plaquette there is an operator $\hat{B}_p$ that commutes with the Hamiltonian. Following Fig.\ref{fig:01_KitaevHoneycombModel_images_Honeycomb_Plaquette}, this operator $\hat{B}_p$ is defined as a product of Pauli matrices around $p$, and is given by  
\beq
	\hat{B}_p = \hat{\sigma}_1^y \hat{\sigma}_2^z \hat{\sigma}_3^x \hat{\sigma}_4^y \hat{\sigma}_5^z \hat{\sigma}_6^x\ .
	\label{eq:ConservedPlaquetteOperators}
\eeq
It is easy to verify that the $\hat{B}_p$ operators commute with the Hamiltonian, which implies that they are conserved quantities, as well as among themselves: 
\beqa
[\hat{H} , \hat{B}_p] &=& 0 ~~~~~ \forall p	\nonumber \\Ê
[\hat{B}_p , \hat{B}_{p'}] &=& 0 ~~~~~ \forall p,p'\ .
\eeqa
Since $\hat{B}_p^2 = 1$, the eigenvalues of these plaquette operators can only take the values $\pm 1$.

\begin{figure}
	\includegraphics[width=.27\textwidth]{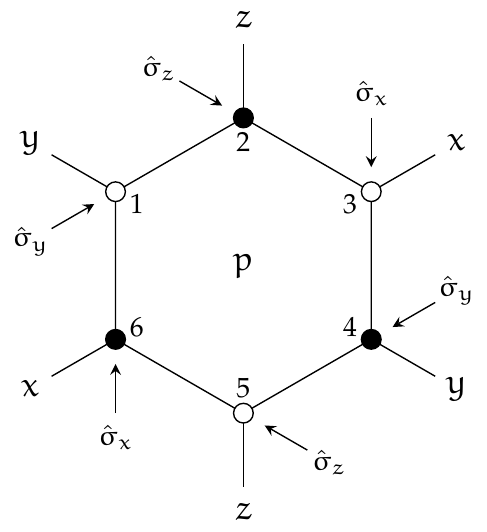}
	\caption{A plaquette of the honeycomb lattice. The conserved quantity $\hat{B}_p$ is constructed as a product of Pauli matrices at every site, where the specific matrix is given by the type of ``outgoing" bond.}
	\label{fig:01_KitaevHoneycombModel_images_Honeycomb_Plaquette}
\end{figure}

\bigskip 
\underline{\emph{(ii) Jordan-Wigner transformation}}

\begin{figure}
	\includegraphics[width=0.4\textwidth]{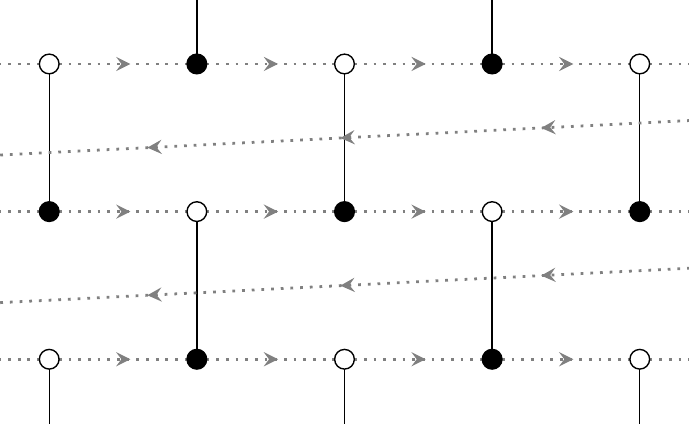}
	\caption{String associated to the ordering of fermionic modes in the Jordan-Wigner transformation of Eq.(\ref{jw}). Boundary terms can be neglected in the thermodynamic limit.}
	\label{fig:JordanWignerPath}
\end{figure}

Let us now label each site as $i,j$, where $i$ is the column and $j$ the row in the brick wall lattice representation. With this notation, we now define the following Jordan-Wigner transformation acting on all the spins of the 2d lattice: 
\beqa
	\hat{\sigma}_{i,j}^{+} &=& 2 \left(  \prod_{j'<j}\prod_{i'}\hat{\sigma}_{i',j'}^z \right)  \left( \prod_{i'<i}\hat{\sigma}_{i',j}^z \right)  \hat{a}_{i,j}^\dagger \nonumber \\
	\hat{\sigma}_{i,j}^- &=& 2 \left( \prod\limits_{j'<j}\prod\limits_{i'}\hat{\sigma}_{i',j'}^z \right) \left( \prod\limits_{i'<i}\hat{\sigma}_{i',j}^z \right) \hat{a}_{i,j}^{\phantom{\dagger}} \nonumber \\
	\hat{\sigma}_{i,j}^z &=& 2 \hat{a}_{i,j}^\dagger \hat{a}_{i,j} - 1 \ . 
	\label{jw}
\eeqa
Here the spin ladder operators are defined in the usual way: $\hat{\sigma}^{\pm} = \left(\hat{\sigma}^x \pm i\hat{\sigma}^y \right)$. The string for the ordering of fermionic modes in the Jordan-Wigner transformation is shown in Fig.\ref{fig:JordanWignerPath}. The different terms in the Hamiltonian of Eq.(\ref{kitaham}) transform then in the following way: 
\beqa
\hat{\sigma}_{i,j}^x \hat{\sigma}_{i+1,j}^x &=& - \left(\hat{a}_{i,j}^\dagger - \hat{a}_{i,j} \right) \left(\hat{a}_{i+1,j}^\dagger + \hat{a}_{i+1,j} \right) \nonumber \\
\hat{\sigma}_{i-1,j}^y \hat{\sigma}_{i,j}^y &=& \left(\hat{a}_{i-1,j}^\dagger + \hat{a}_{i-1,j} \right) \left(\hat{a}_{i,j}^\dagger - \hat{a}_{i,j} \right) \nonumber \\
\hat{\sigma}_{i,j}^z \hat{\sigma}_{i,j+1}^z &=& \left(2 \hat{a}_{i,j}^\dagger \hat{a}_{i,j} -1 \right) \left( 2\hat{a}_{i,j+1}^\dagger \hat{a}_{i,j+1} -1 \right) \ . 
\eeqa

Thus, strings of $\hat{\sigma}^z$'s in the Jordan-Wigner only affect interactions along the $x$ and $y$ bonds, and the resulting fermionic interactions are local, in the sense that they do not have any string of operators attached. After this transformation, the Hamiltonian $H$ reads
\beqa
\hat{H} = &+&J_x \sum\limits_\text{x-links} \left(\hat{a}_{i,j}^\dagger - \hat{a}_{i,j} \right) \left(\hat{a}_{i+1,j}^\dagger + \hat{a}_{i+1,j} \right) \nonumber \\Ê
&-&J_y \sum\limits_\text{y-links} \left(\hat{a}_{i-1,j}^\dagger + \hat{a}_{i-1,j} \right) \left(\hat{a}_{i,j}^\dagger - \hat{a}_{i,j} \right) \nonumber \\
&-&J_z \sum\limits_\text{z-links} \left( 2\hat{a}_{i,j}^\dagger \hat{a}_{i,j} -1 \right) \left( 2\hat{a}_{i,j+1}^\dagger \hat{a}_{i,j+1} -1 \right) 
\eeqa
with $i+j$ even.  The $J_x$ and $J_y$ terms give then quadratic interactions among spinless Dirac fermions. This is good, since quadratic fermionic Hamiltonians can be diagonalized exactly under certain symmetry assumptions such as translation invariance. The $J_z$ term, however, gives some density-density interactions which are quartic in the fermionic modes. We shall see that this quartic term can in fact be further simplified thanks to the conserved quantities $\hat{B}_p$ defined previously. 

\bigskip 
\underline{\emph{(iii) Majorana fermions}}

Let us now introduce two Majorana fermionic operators $\hat{c}_{i,j}$ and $\hat{d}_{i,j}$ at every site of the lattice, such that 
\beq
	\begin{array}{l l l}
		\hat{c}_{i,j} = i\left( \hat{a}_{i,j}^\dagger - \hat{a}_{i,j} \right) & \hat{d}_{i,j} = \hat{a}_{i,j}^\dagger + \hat{a}_{i,j} & i+j  ~ \text{even} \equiv
\begin{tikzpicture} \draw[black,fill=white] (0,0) circle (0.1); \end{tikzpicture} \\
		\hat{c}_{i,j} = \hat{a}_{i,j}^\dagger + \hat{a}_{i,j} & \hat{d}_{i,j} = i\left( \hat{a}_{i,j}^\dagger - \hat{a}_{i,j} \right) & i+j ~ \text{odd} ~ \equiv \begin{tikzpicture} \draw[black,fill=black] (0,0) circle (0.1); \end{tikzpicture} \ . \\Ê
		\end{array} 
\eeq

The original Dirac fermions on each site can be easily reconstructed from these Majorana operators, which obey the usual Majorana relations
\beqa
                &\hat{c}_{i,j}^2 = \hat{d}_{i,j}^2 = 1& \nonumber \\
		&\left\{ \hat{c}_{i,j}, \hat{c}_{i',j'} \right\} = \left\{ \hat{d}_{i,j} , \hat{d}_{i',j'} \right\} = 2\delta_{ii'}\delta_{jj'} & \nonumber \\
		&\left\{ \hat{c}_{i,j} , \hat{d}_{i',j'} \right\} = 0 \ . &
\eeqa

Using these Majorana operators, the Hamiltonian can be rewritten in the following way using pictorial indices that correspond to the even and odd lattice sites respectively: 
\beq
 \hat{H} = -i J_x \sum\limits_\text{x-links} \hat{c}_{\whitesite} \hat{c}_{\blacksite} + i J_y \sum\limits_\text{y-links} \hat{c}_{\blacksite} \hat{c}_{\whitesite} - i J_z \sum\limits_\text{z-links} \left( i\hat{d}_{\blacksite} \hat{d}_{\whitesite} \right)\hat{c}_{\blacksite} \hat{c}_{\whitesite} \ .
\eeq

\begin{figure}
	\includegraphics[width=0.4\textwidth]{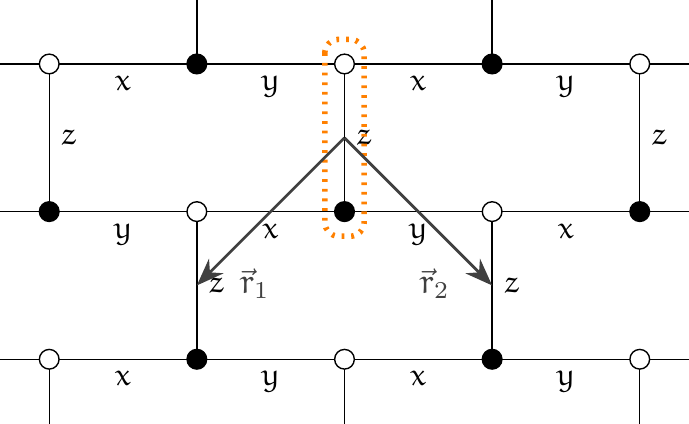}
	\caption{[Color online] Brickwall lattice with unit cell and plane vectors.}
	\label{fig:Honeycomb_Brick_3}
\end{figure}

For the subsequent diagonalization, we first rewrite once again the Hamiltonian using lattice vectors $\vec{r}_1$ and $\vec{r}_2$, and vectors $\vec{r}$ to label the unit cells. This unit cell consists of one even and one odd lattice sites, and vectors $\vec{r}_1$ and $\vec{r}_2$ span then the whole (square) lattice of unit cells, see Fig.(\ref{fig:Honeycomb_Brick_3}).
The sums over all $x$, $y$ and $z$ links in the Hamiltonian are then translated into sums over unit cells, so that we get
\beq
	\hat{H} = i \sum\limits_{\vec{r}} \left(J_x \hat{c}_{\blacksite,\vec{r}}\, \hat{c}_{\whitesite,\vec{r} + \vec{r}_1} + J_y \hat{c}_{\blacksite,\vec{r}}\, \hat{c}_{\whitesite,\vec{r} + \vec{r}_2} - J_z \hat{\alpha}_{\vec{r}}  \hat{c}_{\blacksite,\vec{r}} \, \hat{c}_{\whitesite,\vec{r}} \right)   
	\label{eq:Hamiltonian_UnitCell_ConservedQuantities}
\eeq
with the new operator $\hat{\alpha}_{\vec{r}} = ( i\hat{d}_{\blacksite,\vec{r}}\, \hat{d}_{\whitesite,\vec{r}} )$ defined on each $z$-link of the lattice and being labeled by the unit cell vector $\vec{r}$. The eigenvalues of $\hat{\alpha}_{\vec{r}}$ are good quantum numbers, since these operators commute with the Hamiltonian. This is easy to verify for the first two terms in $\hat{H}$, because the anticommutation relation for single fermions leads to commutations for pairs of fermions. Commutation is also straightforward for the third term and all lattice sites except site $\vec{r}$, for which we have
\beqa
		[ \hat{d}_{\blacksite}\, \hat{d}_{\whitesite}\, \hat{c}_{\blacksite}\, \hat{c}_{\whitesite}\, ,\, \hat{d}_{\blacksite}\, \hat{d}_{\whitesite} ]
		&= &\hat{d}_{\blacksite}\, \hat{d}_{\whitesite} \hat{c}_{\blacksite}\, \hat{c}_{\whitesite}\, \hat{d}_{\blacksite}\, \hat{d}_{\whitesite} - \hat{d}_{\blacksite}\, \hat{d}_{\whitesite}\, \hat{d}_{\blacksite}\, \hat{d}_{\whitesite}\, \hat{c}_{\blacksite}\, \hat{c}_{\whitesite} \nonumber \\
		&=& \hat{d}_{\blacksite}\, \hat{d}_{\whitesite}\, [ \hat{c}_{\blacksite}\, \hat{c}_{\whitesite}\, ,\, \hat{d}_{\blacksite}\, \hat{d}_{\whitesite} ] \nonumber \\
		&=& 0 \ ,
\eeqa
so that commutation holds in all possible cases, and therefore with the full Hamiltonian.

Therefore, operators $\hat{H}$ and $\hat{\alpha}_{\vec{r}}$ can be diagonalized in the same basis. Once a configuration of eigenvalues for all operators $\hat{\alpha}_{\vec{r}}$ is fixed, the model is described by a a Hamiltonian of free Majorana fermions in a static $\mathbb Z_2$ gauge field related to the eigenvalue-configuration of the $\hat{\alpha}_{\vec{r}}$'s. The fact that this generates a static $\mathbb Z_2$ gauge field is because the $\hat{\alpha}_{\vec{r}}$'s are related to the plaquette operators $\hat{B}_p$, which remember, are conserved quantities by themselves. In fact, applying the Jordan-Wigner transformation to $\hat{B}_p$ and rewriting the result with the Majorana fermion operators, one gets
\beq
\hat{B}_p =  \left( i\hat{d}_{\whitesite,1}\, \hat{d}_{\whitesite,3} \right) \left( i\hat{d}_{\blacksite,4}\, \hat{d}_{\blacksite,6} \right) = \hat{\alpha}_{61} \hat{\alpha}_{43} \ , 
\eeq
where the subindices refer to the sites in Fig.(\ref{fig:01_KitaevHoneycombModel_images_Honeycomb_Plaquette}). Therefore, each plaquette $p$ can be regarded as having an independent vortex variable that can take the values $\pm 1$. A vortex configuration for the plaquettes corresponds thus to a given configuration of eigenvalues of the  $\hat{\alpha}_{\vec{r}}$'s. Once this configuration is fixed, the Hamiltonian in this subspace is a free Majorana Hamiltonian that can be diagonalized exactly in the translationally-invariant cases by Fourier-transforming the fermionic modes followed by a Bogoliubov transformation. 

Let us now ``recouple" the Majorana modes into Dirac modes. Since the conserved quantities $\hat{\alpha}_{\vec{r}}$ live on the $z$-links of the lattice, we introduce new Dirac fermions on each $z$-link, defined as follows:
\beqa
	\hat{d}_{\vec{r}} &=& \frac{1}{2} \left( \hat{c}_{\blacksite ,\vec{r}} - i\hat{c}_{\whitesite ,\vec{r}} \right) \nonumber \\
	\hat{d}^\dagger_{\vec{r}}  &=& \frac{1}{2} \left( \hat{c}_{\blacksite ,\vec{r}} + i\hat{c}_{\whitesite ,\vec{r}} \right) \ . 
\eeqa
This yields a model of Dirac fermions on a square lattice with a site-dependent chemical potential, given by
\beqa
\hat{H} = \sum\limits_{\vec{r}} &J_x& \left( \hat{d}^\dagger_{\vec{r}} + \hat{d}_{\vec{r}} \right)\left( \hat{d}^\dagger_{\vec{r} + \vec{r}_1} - \hat{d}_{\vec{r} + \vec{r}_1} \right) \nonumber \\
		+ &J_y& \left( \hat{d}^\dagger_{\vec{r}} + \hat{d}_{\vec{r}} \right)\left( \hat{d}^\dagger_{\vec{r} + \vec{r}_2} - \hat{d}_{\vec{r} + \vec{r}_2} \right) \nonumber \\
	 + &J_z& \alpha_{\vec{r}} \Big( 2\hat{d}^\dagger_{\vec{r}}\hat{d}_{\vec{r}} - 1 \Big)  \ ,
	 \label{fersq}
\eeqa
where we have removed the ``hat" from $\alpha_{\vec{r}}$ to indicate that it is an eigenvalue of operator $\hat{\alpha}_{\vec{r}}$, and therefore we are writing the Hamiltonian in a given flux sector (in fact, this is also what Kitaev calls ``removing hats" \cite{KitaHon}). 

As explained by Kitaev \cite{KitaHon}, a theorem due to Lieb implies that the ground state of the model is in the flux-free sector \cite{Lieb}. This means that, to target the ground state, we need to fix the eigenvalue of $\hat{B}_p$ to $+1$ for all $p$. Choosing the eigenvalues $\alpha_{\vec{r}} = 1$ for all $\vec{r}$ is thus the obvious choice (in fact, all configurations leading to the same vortex sector are equivalent). Notice, though, that other sectors could also be considered if necessary by choosing a different pattern for the quantities $\alpha_{\vec{r}}$. 

\bigskip 
\underline{\emph{(iv) Fourier transformation}}

In the vortex-free sector, the Hamiltonian is translational invariant and can be Fourier-transformed to momentum space in order to be solved. The Fourier transformation for the Dirac fermions is given by
\beqa
\hat{d}_{\vec{r}} &= \frac{1}{\sqrt{N}} \sum\limits_{\vec{k}} \mathrm e^{+i\vec{k} \cdot \vec{r}} \hat{d}_{\vec{k}} \nonumber \\
\hat{d}^\dagger_{\vec{r}} &= \frac{1}{\sqrt{N}} \sum\limits_{\vec{k}} \mathrm e^{-i\vec{k} \cdot \vec{r}} \hat{d}^\dagger_{\vec{k}}	\ .
\eeqa
The resulting Hamiltonian after this transformation is of the BCS-type, and can be written as
\beqa
	&\hat{H} = \sum\limits_{\vec{k}} \left( \varepsilon_{\vec{k}}\hat{d}^\dagger_{\vec{k}}\hat{d}_{\vec{k}} + \frac{1}{2} \left(i\Delta_{\vec{k}}\hat{d}^\dagger_{\vec{k}}\hat{d}^\dagger_{\text{-} \vec{k}}  + h.c. \right) \right) - J_z N& \nonumber \\
	&\varepsilon_{\vec{k}} = 2 \left( J_z - J_x \cos\left( k_1 \right) - J_y \cos\left( k_2 \right) \right)& \nonumber \\
	&\Delta_{\vec{k}} = 2 \left( J_x \sin\left( k_1 \right) + J_y \sin\left( k_2 \right) \right) \ .&
	\label{eq:BCSHamiltonian_1}
\eeqa
Using the symmetric property $\varepsilon_{\vec{k}} = \varepsilon_{\text - \vec{k}}$ we can represent it in the convenient form
\beq
\hat{H} = \sum\limits_{\vec{k}}
		\frac{1}{2}
		\begin{pmatrix}
			\hat{d}_{\vec{k}}^\dagger & \hat{d}_{\text - \vec{k}}
		\end{pmatrix}
		\begin{pmatrix}
			 \varepsilon_{\vec{k}}	& i\Delta_{\vec{k}}	\\
			 -i\Delta_{\vec{k}}^*		& -\varepsilon_{\vec{k}}
		\end{pmatrix}
		\begin{pmatrix}
			\hat{d}_{\vec{k}}	\\
			\hat{d}_{\text - \vec{k}}^\dagger
		\end{pmatrix}\ , 
		\label{eq:BCSHamiltonian_2}
\eeq
where the central $2 \times 2$ matrix is simply $\varepsilon_{\vec{k}} \cdot \sigma^z - \Delta_{\vec{k}} \cdot \sigma^y$ for momentum $\vec{k}$, and both $\varepsilon_{\vec{k}}$ and $\Delta_{\vec{k}}$ are trigonometric polynomials according to Eq.(\ref{eq:BCSHamiltonian_1}). 

\bigskip 
\underline{\emph{(v) Bogoliubov transformation}}

To diagonalize the Hamiltonian in the vortex-free sector we implement now the unitary Bogoliubov transformation
\beqa
\hat{\gamma}_{\vec{k}} &=& u_{\vec{k}} \hat{d}_{\vec{k}} + v_{\vec{k}} \hat{d}_{\text - \vec{k}}^\dagger \nonumber \\
\hat{\gamma}_{\text - \vec{k}}^\dagger &=& - v_{\vec{k}}^{*} \hat{d}_{\vec{k}} + u_{\vec{k}}^{*} \hat{d}_{\text - \vec{k}}^\dagger \ , 
\eeqa
where $u_{\vec{k}}$ and $v_{\vec{k}}$ are complex coefficients satisfying $\vert u_{\vec{k}} \vert^2 + \vert v_{\vec{k}} \vert^2 = 1$. The new Hamiltonian is then
\begin{widetext}
\beq
\hat{H} = \sum\limits_{\vec{k}}
		\frac{1}{2}
		\begin{pmatrix}
			\hat{\gamma}_{\vec{k}}^\dagger & \hat{\gamma}_{\text - \vec{k}}
		\end{pmatrix}
		\begin{pmatrix}
			u_{\vec{k}}	& v_{\vec{k}}	\\
			-v_{\vec{k}}^{*}	& u_{\vec{k}}^{*}	\\
		\end{pmatrix}
		\begin{pmatrix}
			 \varepsilon_{\vec{k}}	& i\Delta_{\vec{k}}	\\
			 -i\Delta_{\vec{k}}^*		& -\varepsilon_{\vec{k}}
		\end{pmatrix}
		\begin{pmatrix}
			u_{\vec{k}}^{*}	& - v_{\vec{k}}	\\
			v_{\vec{k}}^{*}	& u_{\vec{k}}	\\
		\end{pmatrix}
		\begin{pmatrix}
			\hat{\gamma}_{\vec{k}}	\\
			\hat{\gamma}_{\text - \vec{k}}^\dagger
		\end{pmatrix}. 
\eeq
\end{widetext} 
The product of the three $2 \times 2$ matrices above is  
\begin{widetext} 
\beq
	\begin{pmatrix}
		\varepsilon_{\vec{k}} \left( \vert u_{\vec{ k}} \vert^2 - \vert v_{\vec{ k}} \vert^2 \right) + i\Delta_{\vec{ k}} u_{\vec{ k}} v_{\vec{ k}}^{*} - i\Delta_{\vec{ k}} u_{\vec{ k}}^{*} v_{\vec{ k}} & -2\varepsilon_{\vec{ k}} u_{\vec{ k}} v_{\vec{ k}} + i\Delta_{\vec{ k}} u_{\vec{ k}}^2 + i\Delta_{\vec{ k}} v_{\vec{ k}}^2	\\
		-2\varepsilon_{\vec{ k}} u_{\vec{ k}}^{*} v_{\vec{ k}}^{*} - i\Delta_{\vec{ k}} u_{\vec{ k}}^{*2} - i\Delta_{\vec{ k}} v_{\vec{ k}}^{*2} & - \left ( \varepsilon_{\vec{ k}} \left( \vert u_{\vec{ k}} \vert^2 - \vert v_{\vec{ k}} \vert ^2 \right) + i\Delta_{\vec{ k}} u_{\vec{ k}} v_{\vec{ k}}^{*} - i\Delta_{\vec{ k}}u_{\vec{ k}}^{*} v_{\vec{ k}} \right)
	\end{pmatrix} Ê. 
\eeq
\end{widetext} 
The Bogoliubov modes that diagonalize the Hamiltonian are then found by imposing that the non-diagonal terms in the above matrix vanish. For this, it is convenient to express $u_{\vec{k}}$ and $v_{\vec{k}}$ as
\beq
		u_{\vec{k}} = \mathrm e^{i\phi_1}\cos\left( \theta_{\vec{k}}/2 \right)\ , ~~~~~
		v_{\vec{k}} = \mathrm e^{i\phi_2}\sin\left( \theta_{\vec{k}}/2 \right)\ .
\eeq
It is sufficient to use only one condition for vanishing off-diagonal matrix elements, so that we get
\beq
	-2\varepsilon_{\vec{k}} u_{\vec{k}} v_{\vec{k}} + i\Delta_{\vec{k}} u_{\vec{k}}^2 - (-i\Delta_{\vec{k}}) v_{\vec{k}}^2 = 0 
\eeq
and therefore	
\begin{widetext} 
\beq
	\varepsilon_{\vec{ k}} \cos\left( \frac{\theta_{\vec{ k}}}{2} \right) \sin\left( \frac{\theta_{\vec{ k}}}{2} \right) \mathrm e^{i(\phi_1 + \phi_2)} + \Delta_{\vec{ k}} \cos^2\left( \frac{\theta_{\vec{ k}}}{2} \right) \mathrm e^{i(2\phi_1 + \frac{\pi}{2})} -  \Delta_{\vec{ k}} \sin^2\left( \frac{\theta_{\vec{ k}}}{2} \right) \mathrm e^{i(2\phi_2 - \frac{\pi}{2})} = 0 \ , 
	\label{eq:VanishingOffDiagonalTerm}
\eeq
\end{widetext} 
where the relation $\phi_1 + \phi_2 = 2\phi_1 + \pi/2 = 2\phi_2 - \pi/2$ must be fulfilled. Without loss of generality we can choose $\phi_1 = 0$ and $\phi_2 = \pi/2$, so that the condition above can be expressed as
\beq
	\tan\left( \theta_{\vec{k}} \right) = \frac{ \Delta_{\vec{k}} }{\varepsilon_{\vec{k}}}\ .
\eeq
It is now possible to derive the required expressions for the parameters $u_{\vec{ k}}$ and $v_{\vec{ k}}$ using
\beqa
\vert u_{\vec{ k}}\vert^2 - \vert v_{\vec{ k}} \vert^2 = \cos \left( \theta_{\vec{ k}} \right) = \frac{1}{\sqrt{1+\tan^2 \left( \theta_{\vec{ k}} \right)}} = \frac{\varepsilon_{\vec{ k}}}{E_{\vec{ k}}} \nonumber \\
u_{\vec{ k}} v_{\vec{ k}} = \frac i 2 \sin \left( \theta_{\vec{ k}} \right) = \frac i 2 \tan \left( \theta_{\vec{ k}} \right) \cos \left( \theta_{\vec{ k}} \right) = \frac i 2 \frac{\Delta_{\vec{ k}}}{E_{\vec{ k}}} \ , 
	\label{eq:BogoliubovRelations_2}
\eeqa
where $E_{\vec{ k}} = \sqrt{\varepsilon_{\vec{ k}}^2 + \Delta_{\vec{ k}}^2}$ is the quasi-particle excitation energy taken as the positive square root. The parameters then become
\beq
	u_{\vec{ k}} = \sqrt{\frac{1}{2} \left( 1 + \frac{\varepsilon_{\vec{ k}}}{E_{\vec{ k}}} \right)} ~~~
	v_{\vec{ k}} = i\sqrt{\frac{1}{2} \left( 1 - \frac{\varepsilon_{\vec{ k}}}{E_{\vec{ k}}} \right)}\ .
\eeq
Additional properties follow from Eq.(\ref{eq:BogoliubovRelations_2}). In particular, using the relations $E_{\text - \vec{ k}} = E_{\vec{ k}}$ and $\Delta_{\text - \vec{ k}} = - \Delta_{\vec{ k}}$ we see that
\beq
u_{\text - \vec{ k}} v_{\text - \vec{ k}} = - \frac{i}{2} \frac{\Delta_{\vec{ k}}}{E_{\vec{ k}}} \ .
\eeq
Therefore, one of the Bogoliubov coefficients needs to be symmetric while the other one needs to be antisymmetric under $\vec{ k} \rightarrow - \vec{ k}$. Here we take
\beq
	u_{\text - \vec{ k}}= u_{\vec{ k}} ~~~~ v_{\text - \vec{ k}} = - v_{\vec{ k}} \ .
\eeq
This property becomes particularly important for the actual implementation of the Bogoliubov transformation for finite-size systems, which we will use later in this paper for intermediate numerical checks. Finally, the obtained coefficients can be used to express the remaining terms in the Hamiltonian, which now has the diagonal form
\beqa
\hat{H} &=& \sum\limits_{\vec{ k}}
			\frac{1}{2}
			\begin{pmatrix}
				\hat{\gamma}_{\vec{ k}}^\dagger & \hat{\gamma}_{\text - \vec{ k}}
			\end{pmatrix}
			\begin{pmatrix}
				 E_{\vec{ k}}		& 0						\\
				 0						& -E_{\vec{ k}}
			\end{pmatrix}
			\begin{pmatrix}
				\hat{\gamma}_{\vec{ k}}	\\
				\hat{\gamma}^\dagger_{\text - \vec{ k}}
			\end{pmatrix}	\nonumber \\
		&=& \sum\limits_{\vec{ k}} E_{\vec{ k}} \left( \hat{\gamma}_{\vec{ k}}^\dagger \hat{\gamma}_{\vec{ k}} - \frac{1}{2} \right)\ .
\eeqa
The model is then fully diagonalized in the vortex-free sector, and the ground state of the model corresponds to the fermionic vacuum of the Bogoliubov modes.

\subsubsection{Phase diagram} 

\begin{figure}
	\includegraphics[width=.4\textwidth]{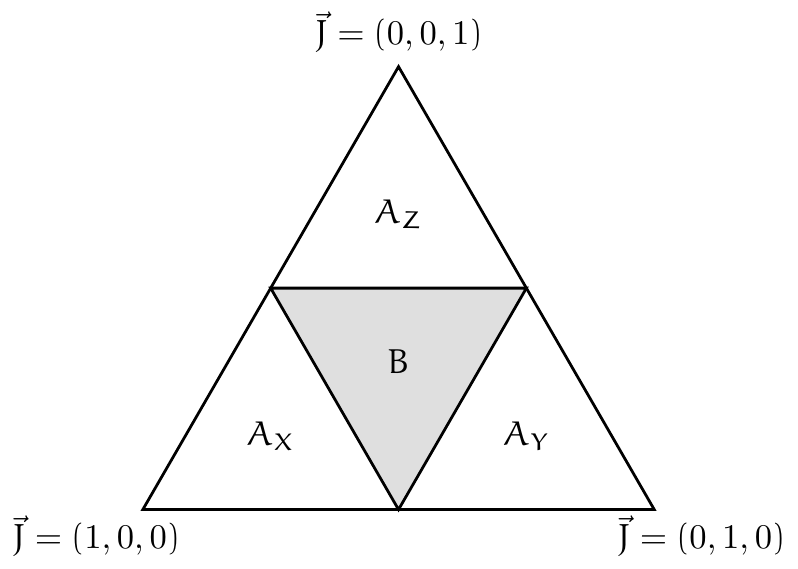}
	\caption{Phase diagram of the Kitaev honeycomb model. A point in the triangle corresponds to relative magnitudes of the spin-spin interactions $J_x$, $J_y$ and $J_z$, which are represented by a normalized vector $\vec{J} = (J_x,J_y,J_z)$ with $J_x + J_y + J_z = 1$. The three different A-phases $A_x, A_y$ and $A_y$ correspond to the case in which one of the couplings predominates over the other two.}
	\label{KitaPD}
\end{figure}

Studying the dispersion relation of the Bogoliubov fermions one can come up with the phase diagram of the model quite straightforwardly, as done originally by Kitaev \cite{KitaHon} and reviewed extensively in many other places (see, e.g., Ref.\cite{BookJiannis} and references therein). Here we will only remind briefly how the phase diagram looks like and what are its most important features. 

The phase diagram of the model is summarized in Fig.\ref{KitaPD}. The energy spectrum is gapless if and only if $|J_x|, |J_y|$ and $|J_z|$ satisfy the triangle inequalities
\beqa
|J_x| &\le& |J_y| + |J_z|, \nonumber \\Ê
|J_y| &\le& |J_x| + |J_z|, \nonumber \\Ê
|J_z| &\le& |J_x| + |J_y|.
\eeqa
In the case of having strict inequalities we are in the B-phase from Fig.\ref{KitaPD}. In this phase there are exactly two Fermi points in momentum space, $\vec{k} = \pm \vec{q}_*$, one in each half of the first Brillouin zone, and corresponding to conic dispersion relations (``Dirac cones"). This means that quasiparticles in this regime are massless and relativistic. In the presence of a small perturbation such as a magnetic field, the phase acquires a gap, has a non-zero Chern number ($=1$), and is chiral and topologically-ordered (time-reversal symmetry being explicitly broken by the perturbation). The gapped quasiparticles are non-abelian Ising anyons. Under certain approximations, the magnetic field perturbation can also be treated in terms of effective three-spin interactions \cite{KitaHon}, which open a gap without destroying the exact solvability of the model. 

If one of the spin-spin interactions predominates over the other two, then the system is in the so-called A-phase, see Fig.\ref{KitaPD}. In this phase the spectrum is gapped, and a perturbation theory analysis taking dimers in one bond direction \cite{KitaHon} shows that it can be approximated by the so-called Toric Code Hamiltonian \cite{toriccode}. The system has therefore non-chiral $\mathbb{Z}_2$ topological order (and thus zero Chern number), and quasiparticles correspond to excitations of the plaquette operators $\hat{B}_p$. The statistics of these quasiparticles corresponds to that of semions, which are abelian anyons. 

\subsection{Tensor networks}

Let us now review briefly the basics of TNs (see, e.g., Ref.\cite{TN} and references therein).

\subsubsection{Basic definitions}

\begin{figure}
\includegraphics[width=0.48\textwidth]{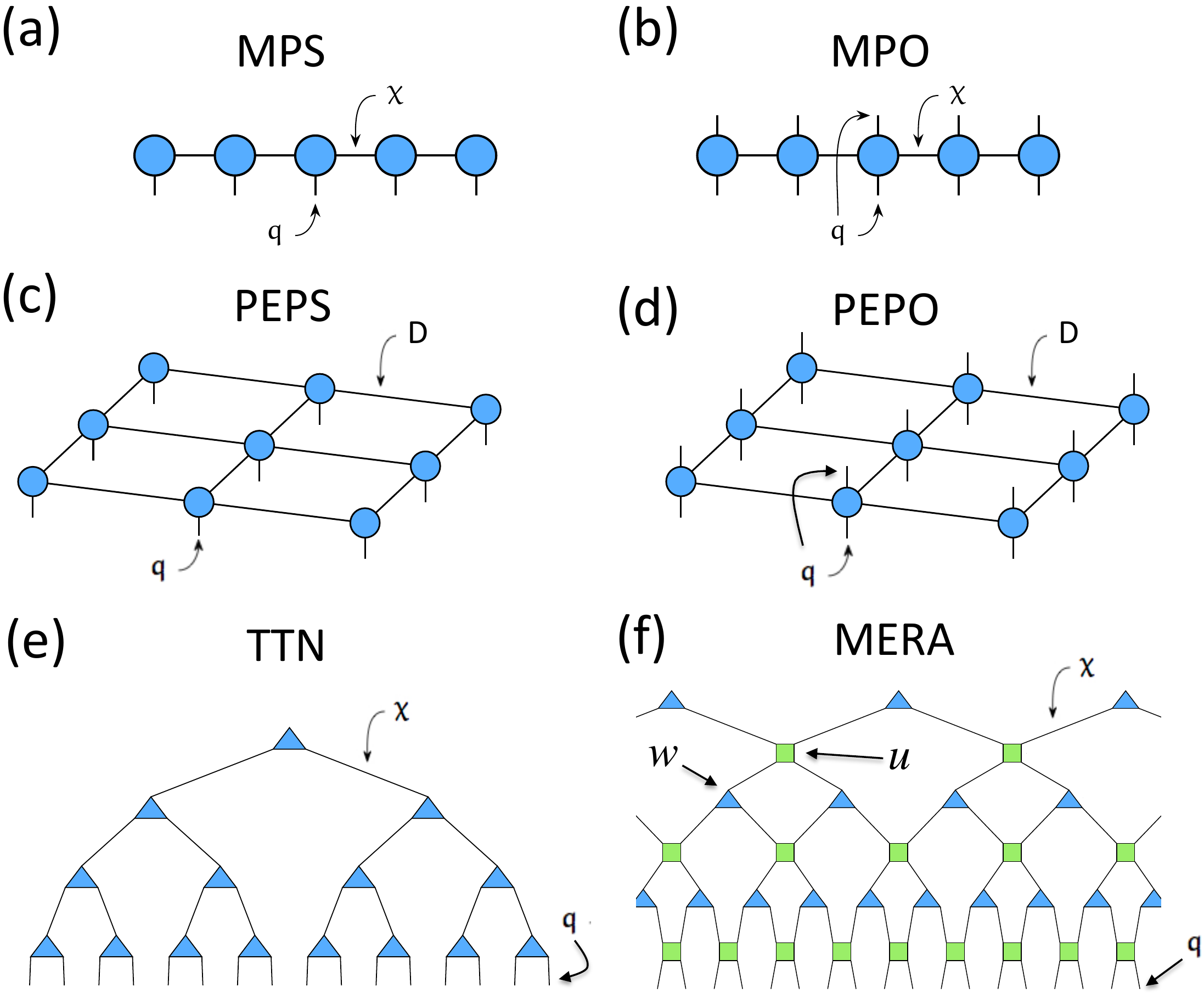}
\caption{[Color online] (a) Matrix Product State (MPS). (b) Matrix Product Operator (MPO). (c) Projected Entangled Pair State (PEPS). (d) Projected Entangled Pair Operator (PEPO). (e) Tree Tensor Network (TTN). (f) 1d binary Multi Scale Entanglement Renormalization Ansatz (MERA), with unitaries $u$ and isommetries $w$. The physical dimension is $q$ in all cases, whereas the bond dimension is $\chi$ for MPS, MPO, TTN and MERA, and $D$ for PEPS and PEPO (though this is a convention). }
\label{fig4}
\end{figure} 

For our purposes, a \emph{tensor} is a multidimensional array of complex numbers. A \emph{Tensor Network} (TN) is a network of tensors whose indices are connected according to some pattern. This connection of indices is done by summing over all the possible values of common indices between tensors. Summing over an index is also called \emph{contracting the index}. Summing over all the possible indices of a given TN is called \emph{contracting the TN}. 

TNs are easily handled by using a diagrammatic notation in terms of \emph{tensor network diagrams}, see, e.g., Fig.(\ref{fig4}). In these diagrams tensors are represented by shapes, and indices in the tensors are represented by lines emerging from the shapes. A TN is thus represented by a set of shapes interconnected by lines. The lines connecting tensors between each other correspond to contracted indices, whereas lines that do not go from one tensor to another correspond to open indices in the TN. 

There are famous examples of TN in the context of many-body physics. For instance, for quantum lattice systems, the classes of MPS and PEPS are suitable to describe quantum states of 1d and 2d systems respectively (see Fig.(\ref{fig4}.a,c)). Other examples make use of an extra ``holographic" dimension accounting for some renormalization group scale, such as Tree Tensor Networks (TTN) \cite{TTN} (Fig.(\ref{fig4}.e)) and the so-called Multi Scale Entanglement Renormalization Ansatz (MERA) (Fig.(\ref{fig4}.f)), which is at the basis of Entanglement Renormalization \cite{MERA}. TNs can also be used to describe operators, such as MPOs in 1d and PEPOs in 2d (Fig.(\ref{fig4}.b,d)). 

\subsubsection{Fermionic tensor networks} 

TNs can be adapted to represent fermionic systems. Several approaches have been developed in order to implement the fermionic statistics at the level of TN algorithms \cite{fmera, fipeps, fPEPS, pit, foc}, in the end being all equivalent (see, e.g., Ref.\cite{revEPJB} for some examples of this). Here we will adopt the approach taken in Ref.\cite{fipeps}, which can be formulated entirely in terms of graphical rules and tensor diagrams. 

\begin{figure}
\begin{centering}
\includegraphics[width=7cm]{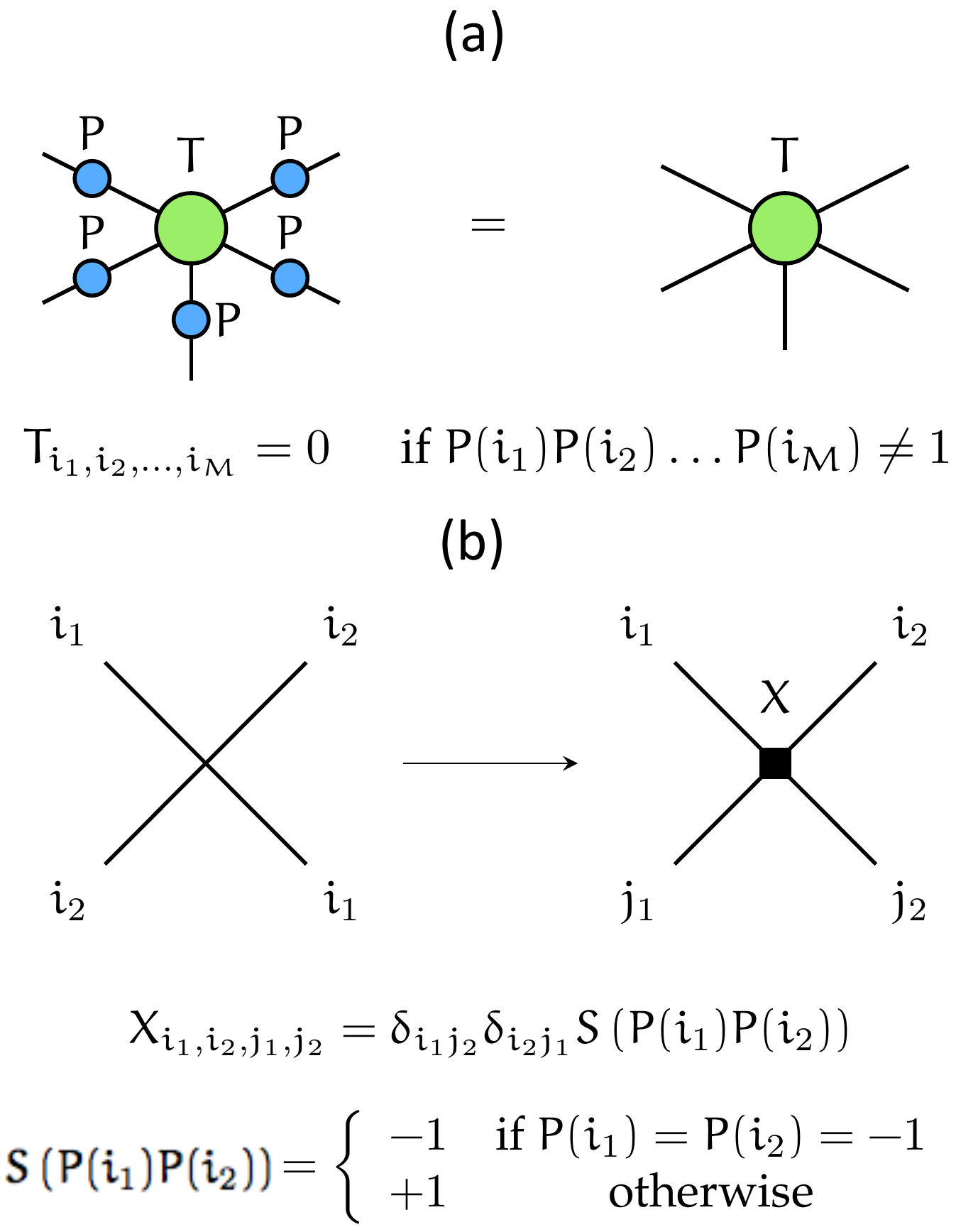} 
\par\end{centering}
\caption{[Color online] TN fermionization rules: (a) parity-invariance of tensors, where $P$ is a representation of the $\mathbb{Z}_2$ parity symmetry operator; (b) crossings are replaced by fermionic swaps.}
\label{Fig10}
\end{figure}

The ``fermionization'' of a TN is based on the following two rules, according to the approach in Ref.\cite{fipeps}: 

\begin{enumerate} 
\item{\emph{Use parity-symmetric tensors}, see Fig.\ref{Fig10}(a). Fermionic parity, i.e., whether the total number of fermions is even or odd, is a good $\mathbb{Z}_2$ symmetry for fermionic systems, and is therefore naturally incorporated in most fermionic gates.}

\item{\emph{Replace line crossings in the planar representation of the TN by fermionic swap gates}, which are defined as in Fig.\ref{Fig10}(b). This replacement has the following physical interpretation: every wire or line in the fermionic TN diagram represents a fermionic degree of freedom. In practice, when an odd number of fermions (odd parity) swap their order with an odd number of fermions, the wavefunction gets multiplied by $-1$. This is what the fermionic swap takes into account, by reading the parity ``charge'' of each index and multiplying by $-1$ whenever appropriate.} 
\end{enumerate}

Following these two simple rules, we can account for diagrammatic representations of TNs for fermionic systems, which as we shall see turns out to be particularly relevant for the Kitaev honeycomb model, since it is mapped to a model of fermions. 

\section{Exact unitary 3d tensor nework}Ê
\label{sec3}

In this section we show how the eigenstates of the Kitaev model, and in particular the ground state, can be expressed as exact 3d unitary TNs. We shall show how this works for relatively small systems and then argue that our construction scales well up to the thermodynamic limit. For the sake of concreteness, to guarantee the correctness of our construction, here we have done intermediate numerical checks for finite-size systems up to  $32$ spins on a honeycomb lattice with periodic boundary conditions. Our approach this time will be ``from the top to the bottom", which means that we will start from a product state of fermionic Bogoliubov modes, and reconstruct the eigenstate step-by-step until we reach the spins on the honeycomb lattice. It could also be done the other way around, starting from spins and ending up in a product state of fermions. However, we believe that the top-bottom approach is more constructive from the TN perspective, and also offers an interesting interpretation of the overall TN as a quantum circuit for a quantum computer building up the many-body wavefunction. We also remark at this point that our TN remains fully fermionic until the mapping to spins done by the Jordan-Wigner transformation. This means that we have to take into account the usual procedures in fermionic TNs, namely, tensors must be parity ($\mathbb{Z}_2$) symmetric, and crossings of wires are accounted for by fermionic SWAP gates \cite{fipeps}.  

\subsection{Fermionic vaccuum and Bogoliubov transformation}

We start in momentum space, where we have a diagonal Hamiltonian and hence unentangled fermionic momentum modes. The possible momenta $\vec{k} = (k_x , k_y)$ for a finite-size $N_x \times N_y$ system are given by
\beq
	k_i = \frac{2\pi n_i}{N_i} ~~~~~~ n_i = -\frac{N_i-1}{2},\hdots,+\frac{N_i-1}{2}\ , 
\eeq
with $i = x,y$ and $N_i$ the size of the system along direction $i$. When taking the thermodynamic limit, the discrete series becomes a continuum of momentum modes. In what follows we describe the procedure for the example of an $8$-spin and $32$-spin honeycomb lattice, which at this stage amount respectively to $2 \times 2$ and $4\times 4$ square lattices of free fermions in momentum space.

\begin{figure}
	\includegraphics[scale=0.7]{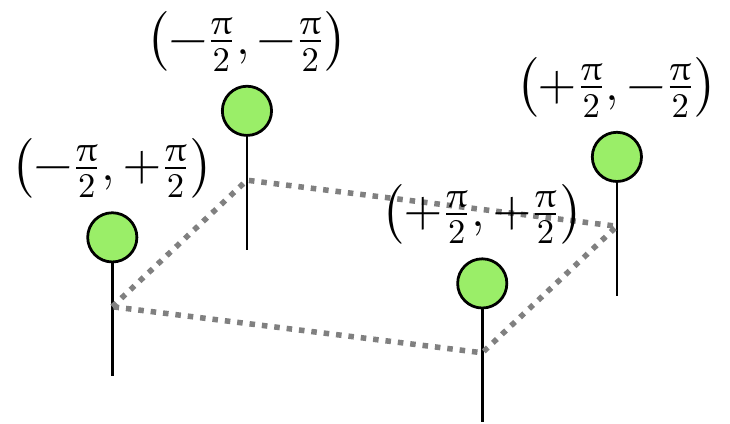}
	\caption{[Color online] Product state in momentum space corresponding to the Bogoliubov vacuum on a $2 \times 2$ square lattice. Dotted line is for reference.}
	\label{fig:TikZ_Fun_HCM_GS_Stage_1}
\end{figure}

The first step to build the TN is the unentangled quantum state in momentum space represented in Fig.(\ref{fig:TikZ_Fun_HCM_GS_Stage_1}). This is a product state of vectors, with physical dimension $2$. The values of the physical index correspond to the fermionic occupation number of Bogoliubov momentum modes. This means that if the physical index is $0$ then the corresponding mode is unoccupied, whereas it is occupied if the value of the physical index is $1$. If we focus on the ground state of the system, then each tensor in the diagram has a non-zero component only for the unoccupied fermionic mode \footnote{Notice also, that in the case of having a non-trivial Fermi surface, then the 2d pattern of occupied/unoccupied modes of the tensors would actually correspond exactly to this Fermi surface.}, or in other words, the overall product state is nothing but the Bogoliubov vacuum. 
 
\begin{figure}
	\includegraphics[width=.5\textwidth]{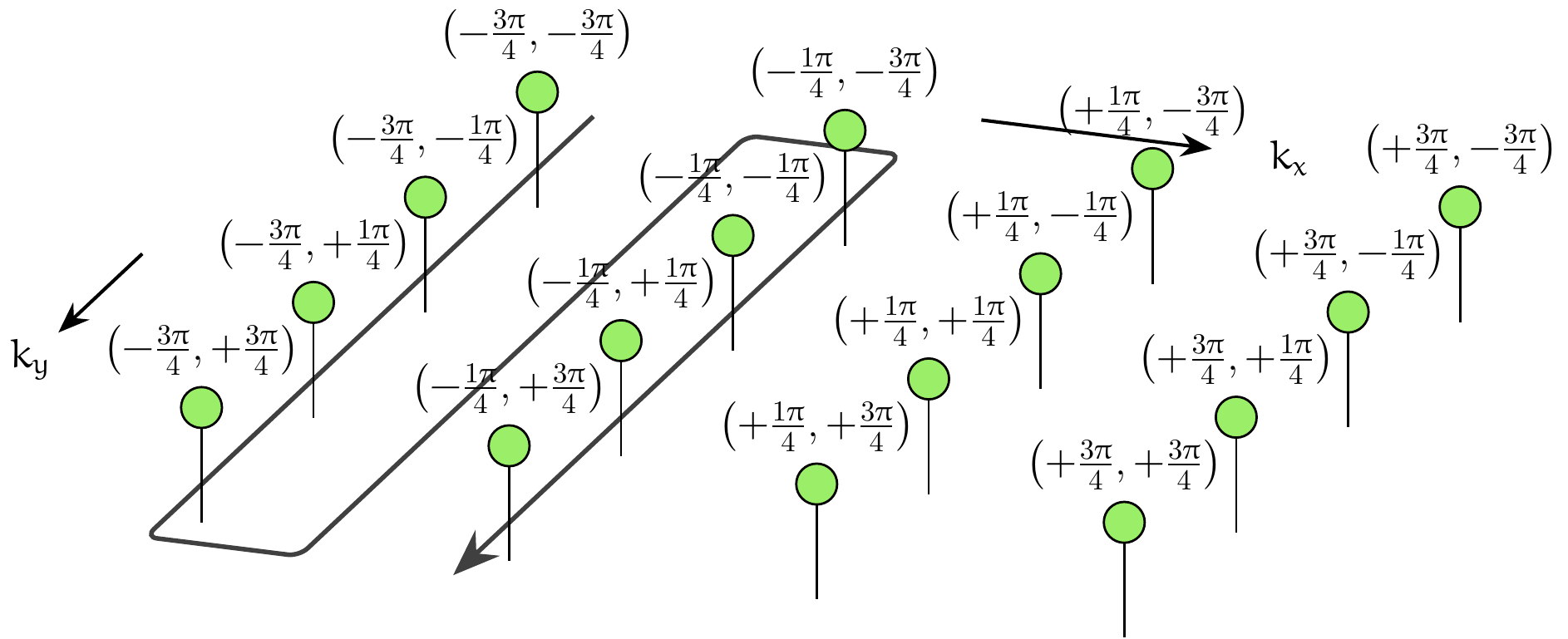}
	\caption{[Color online] Modes along the black arrow are coupled to the ones on the right hand side with opposite momenta, following the ordering indicated by the arrow. The value of $\vec{k}$ for the modes along the arrow also determines the input parameter for the Bogoliubov transformation.}
	\label{fig:TikZ_Fun_Antisymmetry_Momentum}
\end{figure}

In order to couple the modes with opposite momentum we have to undo the Bogoliubov transformation that diagonalized the Hamiltonian in the previous section. Intuitively one can see that this transformation can be built up with 2-body unitary gates. For $N$ fermionic modes we need $N/2$ of such gates. The correct way to perform the Bogoliubov transformation is shown in Fig.(\ref{fig:TikZ_Fun_Antisymmetry_Momentum}) for a $4 \times 4$ lattice, where all the momenta $\vec{k}$ along the black line are the ones entering as gate parameters. The modes along this line are coupled, following the order indicated by the arrow, to the modes with opposite momenta on the right-hand-side of the diagram. Momenta in bigger lattices are also coupled following this prescription.    

The actual Bogoliubov transformation is achieved by a sequence of $2$-body unitary gates acting on pairs of fermions with opposite momentum. These gates are described by the unitary matrix
\beq
	\hat{B}_{\vec{k}} =
	\begin{pmatrix}
		u_{\vec{k}}		& 0		& 0		& v_{\vec{k}} \\
		0		& 1		& 0		& 0		\\
		0		& 0		& 1		& 0		\\
		-v_{\vec{k}}^{*}		& 0		& 0		& u_{\vec{k}}^{*}
	\end{pmatrix} \ , 
\eeq
where rows/columns are in the $\{ \ket{00}, \ket{01}, \ket{10}, \ket{11} \}$ occupation basis of the respective modes. Besides the momentum dependency, both $u_{\vec{k}}$ and $v_{\vec{k}}$ depend non-trivially on $J_x$, $J_y$ and $J_z$. Also, due to intrinsic properties of the Bogoliubov transformation, the coefficients must satisfy the condition
\beq
	u_{- \vec{k}} = u_{\vec{k}} ~~~~~~ v_{- \vec{k}} = -v_{\vec{k}}\ .
\eeq
On top of this we have also observed that, in order to generate the correct transformations, the sign of $v_{\vec{k}}$ needs to be chosen appropriately and in accordance with the values of $J_x$ and $J_y$. A numerical check indicates that the correct convention is the one in Table \ref{tab:BogoliubovCoefficientSign}, where the conditions for a negative overall sign of $v_{\vec{k}}$ are specified.
\begin{table}[h]
	\centering
	\begin{tabular}{||c||c||c||} 
	\hline 
	  ~~Condition 1	~~    				& ~~ Condition 2	~~	& ~~Condition 3~ \\ \hline \hline 
								& $J_x \ge 0$, $J_y \ge 0$ 	& $k_x  <  0$		\\
		{$|J_x| \ge |J_y|$}			& $J_x \ge 0$, $J_y  <  0$	 & $k_x  <  0$		\\
								& $J_x  <  0$, $J_y \ge 0$	 & $k_x \ge 0$		\\
								& $J_x  <  0$, $J_y  <  0$		& $k_x \ge 0$		\\ \hline 
								& $J_x \ge 0$, $J_y \ge 0$ 	& $k_y  <  0$		\\
		{$|J_x|  < |J_y|$}			& $J_x \ge 0$, $J_y  <  0$		& $k_y \ge 0$		\\
								& $J_x  <  0$, $J_y \ge 0$		& $k_y  <  0$		\\
							        & $J_x  <  0$, $J_y  <  0$		& $k_y \ge 0$		\\ \hline 														
	\end{tabular}
	\caption{Conditions for $\text{sign}(v_{\vec{k}}) = -1$ in the Bogoliubov transformation.}
	\label{tab:BogoliubovCoefficientSign}
\end{table}

\begin{figure}
    \includegraphics[scale=0.7]{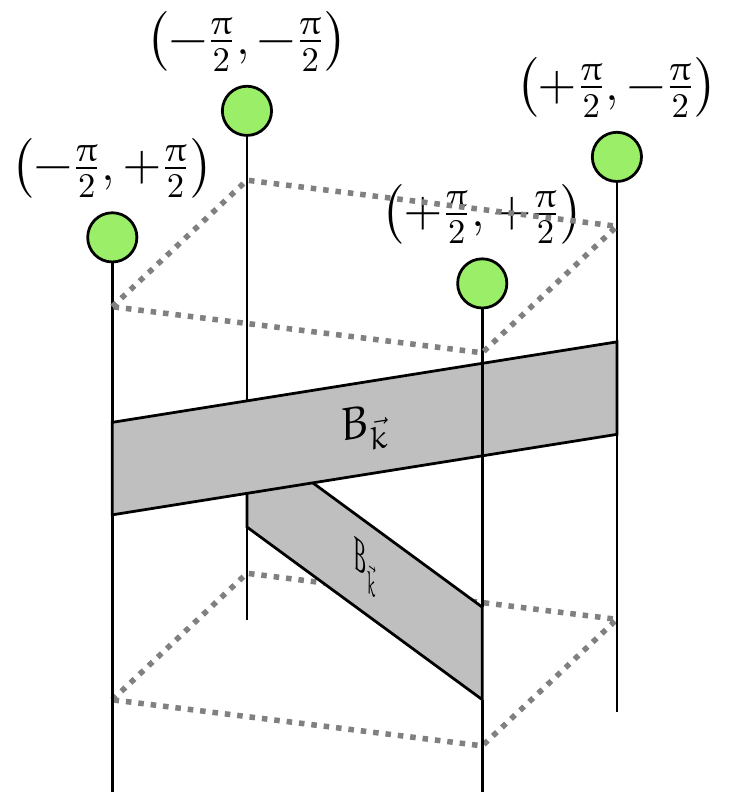}
  \caption{[Color online] Entangled state for the $2 \times 2$ square lattice, obtained after applying the Bogoliubov transformation. The construction can be easily scaled up to the thermodynamic limit using the prescriptions described in the text. This is an entangled state of fermionic momentum modes. Dotted lines are for reference.}
  \label{fig:TikZ_Fun_HCM_GS_Stage_2}
\end{figure}

This Bogoliubov transformation works smoothly for square lattices with an even number of sites in both directions, which can be conveniently scaled up to the thermodynamic limit \footnote{For rectangular lattices, we have also seen that the conventions depending on the parameter regimes are different to the ones presented here for square lattices.}. Applying this transformation to the Bogoliubov vaccuum we get the entangled state in momentum space shown in Fig.\ref{fig:TikZ_Fun_HCM_GS_Stage_2}.

\begin{figure}
    \includegraphics[width=0.25\textwidth]{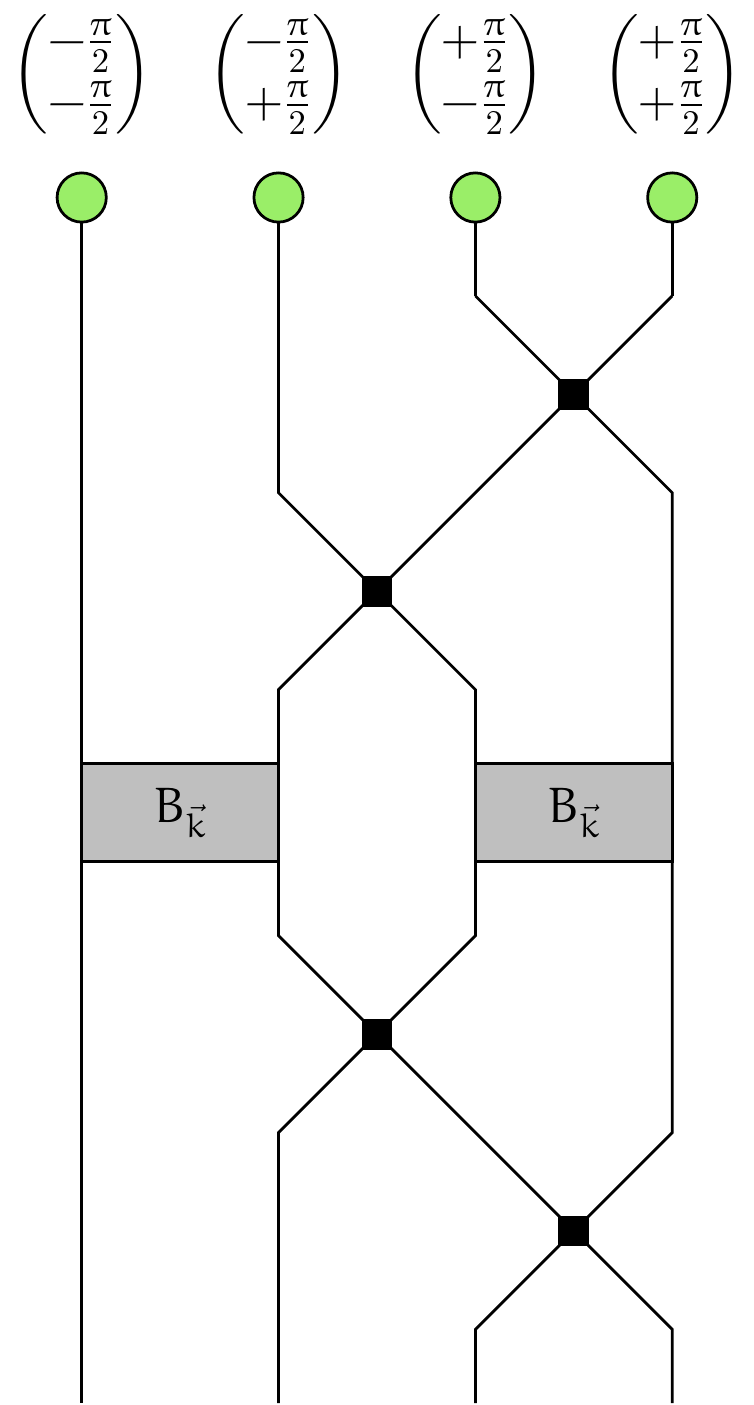}
  \caption{[Color online] A possible planar projection of the TN in Fig.\ref{fig:TikZ_Fun_HCM_GS_Stage_2}, accounting for a specific fermionic order in second quantization (from left to right). To satisfy the fermionic anti-commutation relations, fermionic SWAP gates (black squares) are introduced each time two fermions are swapped.}
  \label{fig:TikZ_Fun_HCM_GS_Stage_3}
\end{figure}

It is clear now, that the TN we are constructing will have a 3d structure. The TN diagrams will therefore be projections of this network in the 2d plane of the paper. This is trivial for bosons, but for fermions one needs to take into account that different projections will produce different patterns of fermionic SWAP gates, which in turn correspond to different orderings of fermionic modes in second quantization \cite{fipeps}. In our case, the resulting planar structure for the TN in Fig.(\ref{fig:TikZ_Fun_HCM_GS_Stage_2}) is shown in Fig.(\ref{fig:TikZ_Fun_HCM_GS_Stage_3}), where the black squares correspond to fermionic SWAP tensors. Such tensors account for the fermionic anticonmutation relation: whenever two odd-parity sectors are swapped, we get a multiplying factor of $-1$. We verified numerically that this construction correctly produces all the eigenstates of the intermediate fermionic Hamiltonian in Eq.(\ref{eq:BCSHamiltonian_2}) for lattices up to $4 \times 4$ sites, and it can be easily generalized to larger lattices and to the thermodynamic limit.

\subsection{Fourier transformation and spectral TN}Ê

The next step in our TN construction is to undo the Fourier transformation of the fermions, bringing the modes from momentum space back to real space. Here we assume, for convenience of the Fourier transformation, that the number of modes is a power of two. 

The TN needed for this step was originally proposed in Ref.\cite{specTN}. In that reference, the so-called \emph{spectral TN} was put forward in order to represent the Fourier transformation of fermions. In our case, this is exactly the piece we need at this step, and we review it briefly in what follows. 

\begin{figure}
    \includegraphics[width=.4\textwidth]{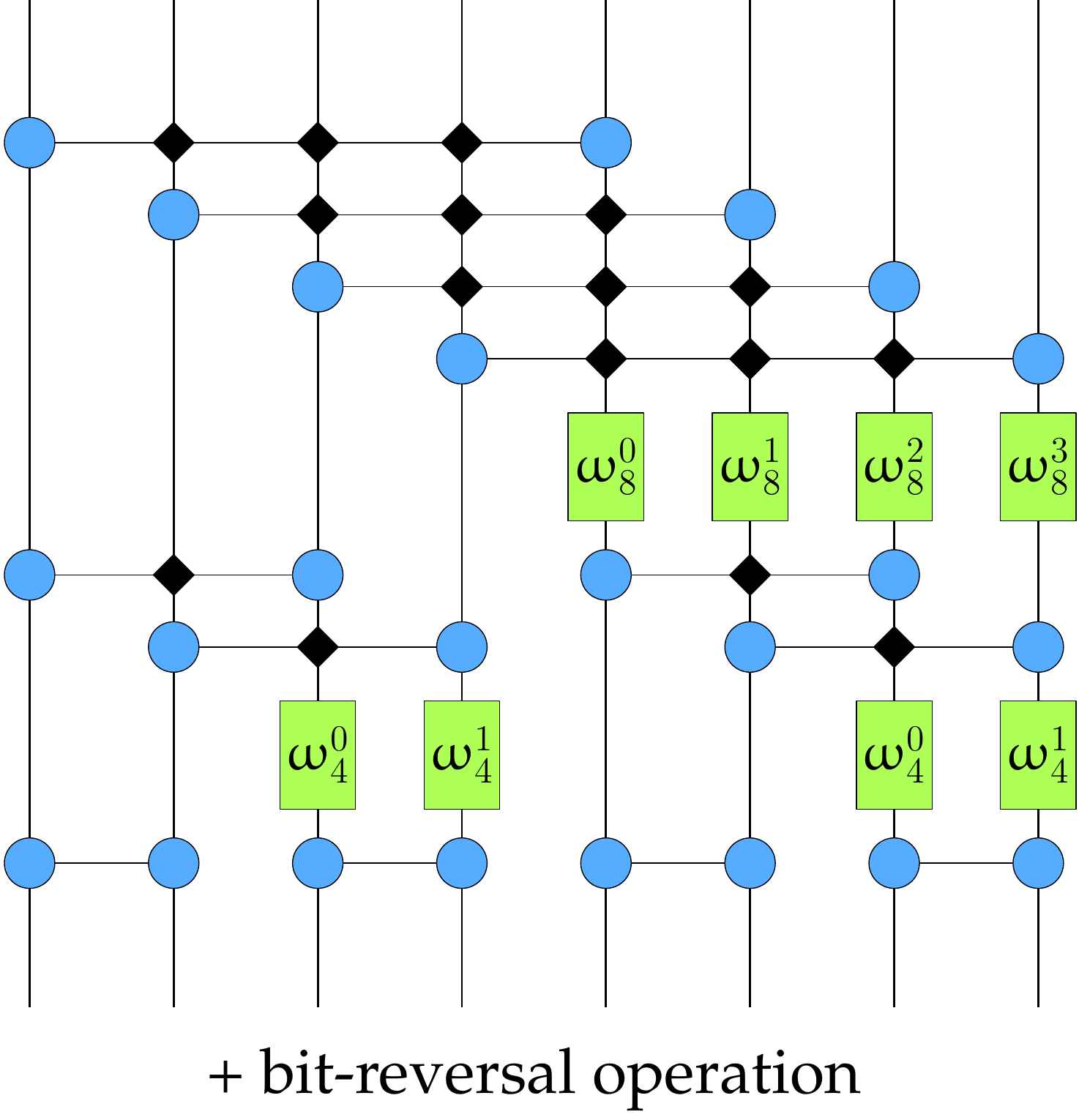}
  \caption{[Color online] Quantum circuit, or spectral TN, for the Fourier transformation for eight fermionic modes, implemented entirely in terms of one- and two-body gates. The upper side of the diagram corresponds to momentum space, and the lower part to real space. Crossings of wires correspond to fermionic SWAP gates.}
  \label{fig:TikZ_Fun_FourierNetwork}
\end{figure}

The quantum circuit of the Fourier transformation translates the quantum state between the momentum and real spaces, and can be  decomposed into a series of sparse operations. More precisely, the fermionic Fourier transformation over $N$ sites (no matter the lattice) can be decomposed into two parallel transformations, each one on $N/2$ sites. To see this notice that, e.g., the creation operator $\hat{c}_k^\dagger$ in 1d ($k$ being the momentum variable) is given in general by
\beqa
	\hat{c}_k^\dagger &=& \frac{1}{\sqrt{N}} \sum\limits_{x=0}^{N-1} \mathrm e^{\frac{2\pi i k x}{N}} \hat{c}_x^\dagger \nonumber \\ 
	&=& \frac{1}{\sqrt{N}} \sum\limits_{x=0}^{N/2-1} \mathrm e^{\frac{2\pi i k x}{N/2}} \hat{c}_{2x}^\dagger \nonumber \\Ê
	&+& \frac{1}{\sqrt{N}} \mathrm e^{\frac{2\pi i k}{N}} \sum\limits_{x=0}^{N/2-1} \mathrm e^{\frac{2\pi i k x}{N/2}} \hat{c}_{2x+1}^\dagger \ ,
\eeqa
where in the second equation we have splited the sum into one transformation for even sites and one transformation for odd sites $x$. This implies, as explained in Ref.\cite{specTN}, that the whole transformation can be implemented entirely by one- and two-body gates. The trick to achieve this is clear: simply keep splitting the sum into halves, until ending up eventually with two-site transformations and one-body gates accounting for the relative phases. In Fig.(\ref{fig:TikZ_Fun_FourierNetwork}) we show this decomposition for the example of 8 sites. The permutation at the bottom of the circuit is a ``bit-reversal operation" that restores the correct order of the fermionic modes. Notice, that since this is also a fermionic TN, all crossings of lines in the TN diagram need to be accounted for by fermionic SWAP gates. 

The entangling two-body gate depicted in Fig.(\ref{fig:TikZ_Fun_FourierNetwork}) is the essential tensor here, and corresponds to  the Fourier transformation of two fermionic modes. In the occupation number basis $\{ \ket{00},\ket{01},\ket{10},\ket{11} \}$ of the two modes, the action of the gate is given by the $4 \times 4$ unitary matrix
\beq
	\hat{F}_2 = \begin{pmatrix}
		1			& 0						& 0 					& 0	\\
		0			& \frac{1}{\sqrt 2}		& \frac{1}{\sqrt 2}		& 0	\\
		0			& \frac{1}{\sqrt 2}		& -\frac{1}{\sqrt 2}	& 0	\\
		0			& 0						& 0						& -1
	\end{pmatrix} 
\eeq
where the last entry accounts for the anti-commutation of two occupied fermionic modes being swapped. As expected, this matrix is parity-preserving. The $4 \times 4$ matrix can easily be rewritten by a 2-site fermionic Matrix Product Operator (MPO) as in Fig.(\ref{fig:TikZ_Fun_FourierNetwork}) with bond dimension $4$ \footnote{This can be done in many ways, e.g., via singular value decomposition.}. The one-body gates include the correct twiddle factor $\mathrm \exp(2\pi i k/N)$ and correspond to 
\beq
\hat\omega_N^k = \left( \hat{\sigma}^z \right)^{\frac{2k}{N}} \, 
\eeq
with $\hat{\sigma}^z$ the usual z-Pauli matrix. 

\begin{figure}
	\centering
	\includegraphics[width=.4\textwidth]{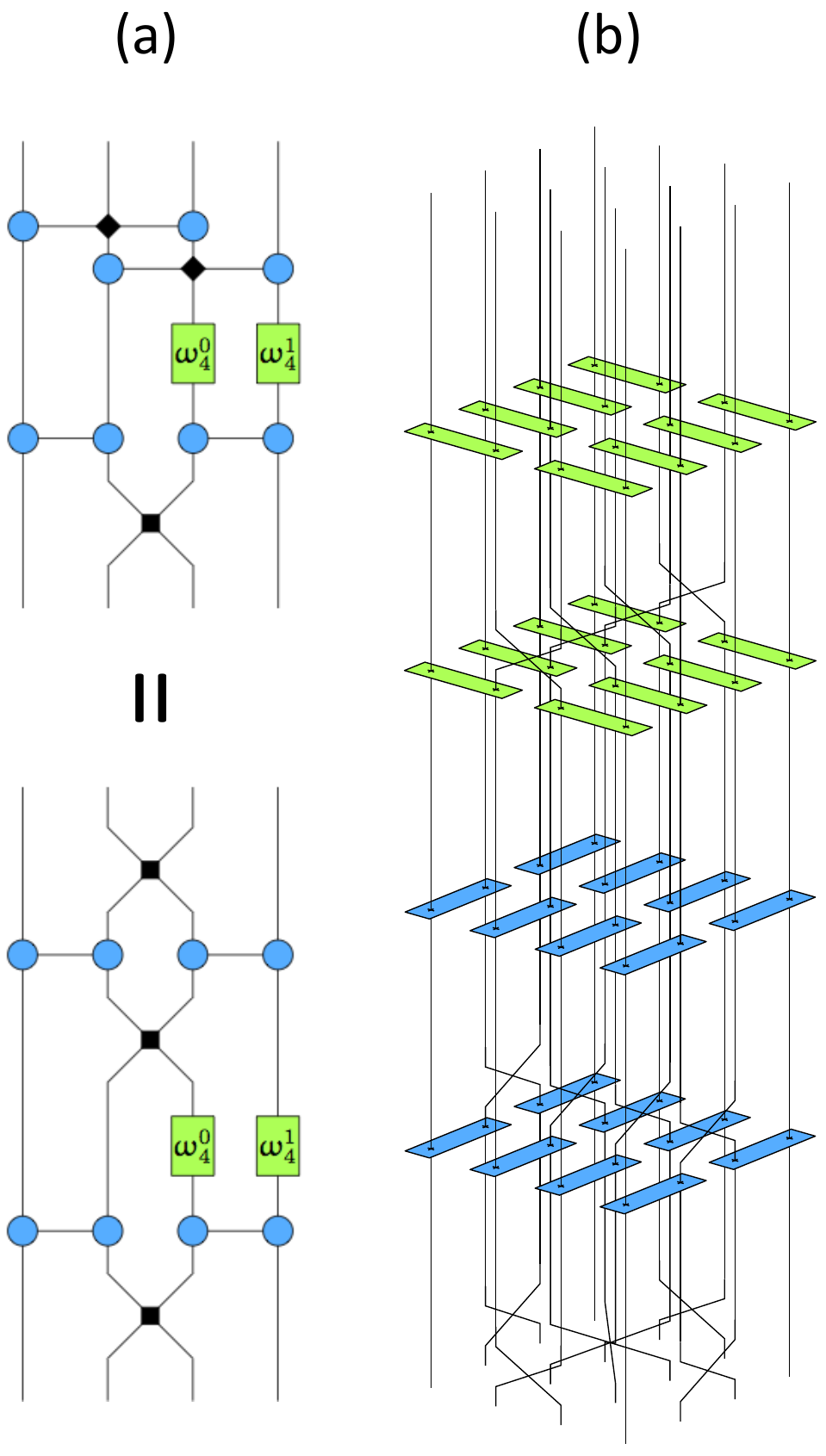}
	\caption{[Color online] (a) TN for the four-site Fourier transformation with a bit-reversal permutation at the end. This 1d structure is repeated along $x$ and $y$ directions to achieve a two-dimensional Fourier transformationation, such as the one in (b) for a $4 \times 4$ square lattice. In (b), one-body gates and two-site MPOs are contracted into two-body gates (for simplicity of the diagram), a bit-reversal permutation is performed at the bottom, and fermionic SWAPS are not explicitly highlighted also for simplicity.}
	\label{fig:TikZ_Fun_FourierNetwork_4x4Sites}
\end{figure}

Moreover, this procedure can be easily extended to higher-dimensional systems \cite{specTN} by applying sequentially the Fourier transformation along each direction, and in parallel for every row or column. To give a concrete example of this, we show the Fourier transformation for a $4 \times 4$ square lattice in Fig.(\ref{fig:TikZ_Fun_FourierNetwork_4x4Sites}), where the quantum circuit in Fig.(\ref{fig:TikZ_Fun_FourierNetwork_4x4Sites}.a) is repeated four times along the $x$ and $y$ directions. For the sake of simplicity in the diagram, in Fig.(\ref{fig:TikZ_Fun_FourierNetwork_4x4Sites}.b) we contracted the one-body gates and the two-site MPOs into some two-body gates, and did not highlight fermionic SWAPS explicitly. 

Finally, we have verified numerically that the state constructed in this way up to this point (concatenating the Bogoliubov TN with the spectral TN) is the exact eigenstate of the intermediate fermionic Hamiltonian in real space given in Eq.(\ref{fersq}) for a $4 \times 4$ lattice. Some subtilities coming from the finite-size of the lattice are however in order, which we explain in Appendix \ref{appA}. 

\subsection{Majorana braiding TN}Ê

A crucial step in the construction of the TN comes at this point. Now, the Dirac fermionic modes in real space are splitted into Majorana modes. The same is done for the fermionic Dirac modes providing the vortex sector. And then, the Majoranas from the real space Dirac fermions are recombined with the Majoranas from the vortex Dirac fermions, into new Dirac modes living on the sites of a  square lattice. Algebraically this step is clear, as we reviewed (in reverse) in Sec.\ref{sec2}. However, at the TN level, this leads in principle to a non-trivial construction. The first problem comes from the fact that it is, actually, not possible to associate a Fock space to Majorana fermions. How are we then going to represent them using TNs? The second problem is that we should be able to represent the splitting of Dirac fermions into Majorana modes, as well as the recombination of Majorana modes into Dirac fermions. The third issue, which is more subtle, is that all these splittings-and-recombinations lead necessarily to \emph{braidings of Majorana fermions}, and we know that they behave like non-abelian anyons! The key result of this section is that, in fact, it is actually possible to  put all these ingredients together into a very nice and natural TN description, as we shall describe in what follows. 

\subsubsection{Majoranas as non-abelian anyons}

Let us start by reviewing the answer to the following question: ``what does it mean that Majorana modes behave like non-abelian anyons?". For this, we consider two spinless Dirac fermions with creation operators $\hat{f}_1^\dagger$ and $\hat{f}_2^\dagger$. Creation and annihilation operators for both fermions can be rewritten in terms of Majorana fermions $\hat{\gamma}_1, \hat{\gamma}_2, \hat{\gamma}_3$ and $\hat{\gamma}_4$ as
\beqa
	\hat{f}_{1} = \frac{1}{2} \left( \hat{\gamma}_2 - i\hat{\gamma}_1 \right) &~~~~~& \hat{f}_{1}^\dagger = \frac{1}{2} \left( \hat{\gamma}_2 + i\hat{\gamma}_1 \right) \nonumber \\
	\hat{f}_{2} = \frac{1}{2} \left( \hat{\gamma}_3 - i\hat{\gamma}_4 \right)  &~~~~~& \hat{f}_{2}^\dagger = \frac{1}{2} \left( \hat{\gamma}_3 + i\hat{\gamma}_4 \right) \ .
\eeqa
Complementary, the Majorana fermions correspond to the real and imaginary part of the original Dirac fermions, since 
\beqa
	\hat{\gamma}_1 &=&-i \left(\hat{f}_{1}^\dagger - \hat{f}_{1}\right) ~~~~~~ \hat{\gamma}_2 = \hat{f}_{1}^\dagger + \hat{f}_{1} \nonumber \\
	\hat{\gamma}_3 &=& \hat{f}_{2}^\dagger + \hat{f}_{2} ~~~~~~~~~~~~~ \hat{\gamma}_4 = -i\left( \hat{f}_{2}^\dagger - \hat{f}_{2} \right)\ .
\eeqa
Therefore $\hat{\gamma}_i = \hat{\gamma}_i^\dagger$ and also $\{ \hat{\gamma}_i, \hat{\gamma}_j \} = 2 \delta_{ij}$, which follows easily from the anticonmutation relation for Dirac fermions. Moreover, one has that $\hat{\gamma}_i^2 = 1$. This means, in particular, that one can not even define a number operator $\hat{n}_i = \hat{\gamma}_i^\dagger \hat{\gamma}_i$, so that an occupation number representation is not appropriate. Notice also that, given a Dirac mode, there is a intrinsic gauge degree of freedom in the definition of its Majorana components since these could be, e.g., multiplied by some relative phase. The convention (gauge) that we used above is the one that will be useful later for our TN construction \footnote{More details on the practical issues of building TNs with Majorana fermions (and beyond) will appear in a forthcoming publication.}. 

It is possible to see \cite{MajBrad}, that the clockwise and anticlockwise braidings (or swaps) of two Majorana fermions $\hat{\gamma}_i$ and $\hat{\gamma}_j$ can be accounted for respectively by the operators
\beqa
	\hat{B} &=& \frac{1}{\sqrt{2}} \left( 1 + \hat{\gamma}_i \hat{\gamma}_j \right) ~~~~\text{(clockwise)} \nonumber \\Ê
        \hat{\bar{B}} &=& \frac{1}{\sqrt{2}} \left( 1 - \hat{\gamma}_i \hat{\gamma}_j \right) ~~~~\text{(anticlockwise)} \ .
\eeqa
If $\hat{\gamma}_i$ and $\hat{\gamma}_j$ are the Majorana modes of one Dirac fermion, then the above operators will be a one-body gate in the Fock space of the Dirac fermion, and will do an ``internal" braiding within the Dirac mode. However, if $\hat{\gamma}_i$ and $\hat{\gamma}_j$ are Majorana modes of different Dirac fermions, then $\hat{B}$ and $\hat{\bar{B}}$ will be two-body operators acting on the Fock space of the two Dirac fermions, and will implement an ``external" braiding, effectively exchanging one of the Majorana components inside of each Dirac fermion. The effect of such operations can be easily computed by rewriting them in terms of the original Dirac modes, to which we know we can associate a Fock space. Let us be more precise about this. For the case of a braiding of the modes $\hat{\gamma}_3$ and $\hat{\gamma}_4$ inside $\hat{f}_2$, one has the following for the clockwise and anticlockwise cases: 
\beqa
	\hat{B} &=& \frac{1}{\sqrt 2} \left( 1 + i(\hat{f}_2^\dagger \hat{f}_2 - \hat{f}_2 \hat{f}_2^\dagger) \right) \nonumber \\Ê
	\hat{\bar{B}} &=& \frac{1}{\sqrt 2} \left( 1 - i(\hat{f}_2^\dagger \hat{f}_2 - \hat{f}_2 \hat{f}_2^\dagger) \right). ~~
\eeqa	
It is then very easy to compute the matrix elements of these operators in the basis $\{ \ket{0}, \ket{1} \}$ of the Dirac fermion, with $\ket{0} = (0, 1)^t$ and $\ket{0} = (1, 0)^t$. These are given by the $2 \times 2$ unitary matrices
\beq
	\hat{B} = \frac{1}{\sqrt 2}
	\begin{pmatrix}
			1+i		& 0	\\
			0		& 1-i
	\end{pmatrix} ~~
	\hat{\bar{B}} = \frac{1}{\sqrt 2}
	\begin{pmatrix}
			1-i		& 0	\\
			0		& 1+i
	\end{pmatrix} 
\eeq
where the first row/column is for the state $\ket{0}$, and the second for $\ket{1}$. Additionally for the case of a braiding of the modes $\hat{\gamma}_2$ from $\hat{f}_1$ and $\hat{\gamma}_3$ from $\hat{f}_2$, the corresponding braiding operators can be rewritten in terms of the Dirac modes as follows: 
\beqa
	\hat{B} &=& \frac{1}{\sqrt 2} \left( 1 - i(\hat{f}_1^\dagger \hat{f}_2^\dagger - \hat{f}_1^\dagger \hat{f}_2 + \hat{f}_1 \hat{f}_2^\dagger - \hat{f}_1 \hat{f}_2) \right) \nonumber \\Ê
	\hat{\bar{B}} &=& \frac{1}{\sqrt 2} \left( 1 + i(\hat{f}_1^\dagger \hat{f}_2^\dagger - \hat{f}_1^\dagger \hat{f}_2 + \hat{f}_1 \hat{f}_2^\dagger - \hat{f}_1 \hat{f}_2) \right) \ . 
\eeqa
These are clearly two-body fermionic gates acting on the Fock spaces of the Dirac fermions $\hat{f}_1$ and $\hat{f}_2$. On the two-particle basis $\{ \ket{00}, \ket{01}, \ket{10}, \ket{11} \}$, the matrix elements are given by the $4 \times 4$ matrices 
\beq
	\hat{B} = \frac{1}{\sqrt 2}
	\begin{pmatrix}
		1		& 0		& 0		& -i	\\
		0		& 1		& i	& 0		\\
		0		& -i	& 1		& 0		\\
		i		& 0		& 0		& 1
	\end{pmatrix}~	
	\hat{\bar{B}} = \frac{1}{\sqrt 2}
	\begin{pmatrix}
		1		& 0		& 0		& i	\\
		0		& 1		& -i	& 0		\\
		0		& i	& 1		& 0		\\
		-i		& 0		& 0		& 1
	\end{pmatrix}, 
\eeq
where the first row/column corresponds to $\ket{00}$, the second to $\ket{01}$, and so on. Notice that, as expected, all these operators preserve the fermionic parity, and have therefore a $\mathbb{Z}_2$ symmetry. Finally, notice that other swaps, e.g., between $\hat{\gamma}_1$ and $\hat{\gamma}_4$, would follow easily by concatenating the ones that we just described.  

\subsubsection{Building up a TN of Majorana braidings}

We are now in position to analyse how the different operations with Majorana fermions in the Kitaev honeycomb model can be accounted for in the TN picture. The key point to do this, is to realize that such operations can always be represented by unitary operators in the Fock space of the Dirac fermions, as we have just shown. Therefore, it is possible to represent them using the usual fermionic TN language. 

\begin{figure}
    \includegraphics[width=.47\textwidth]{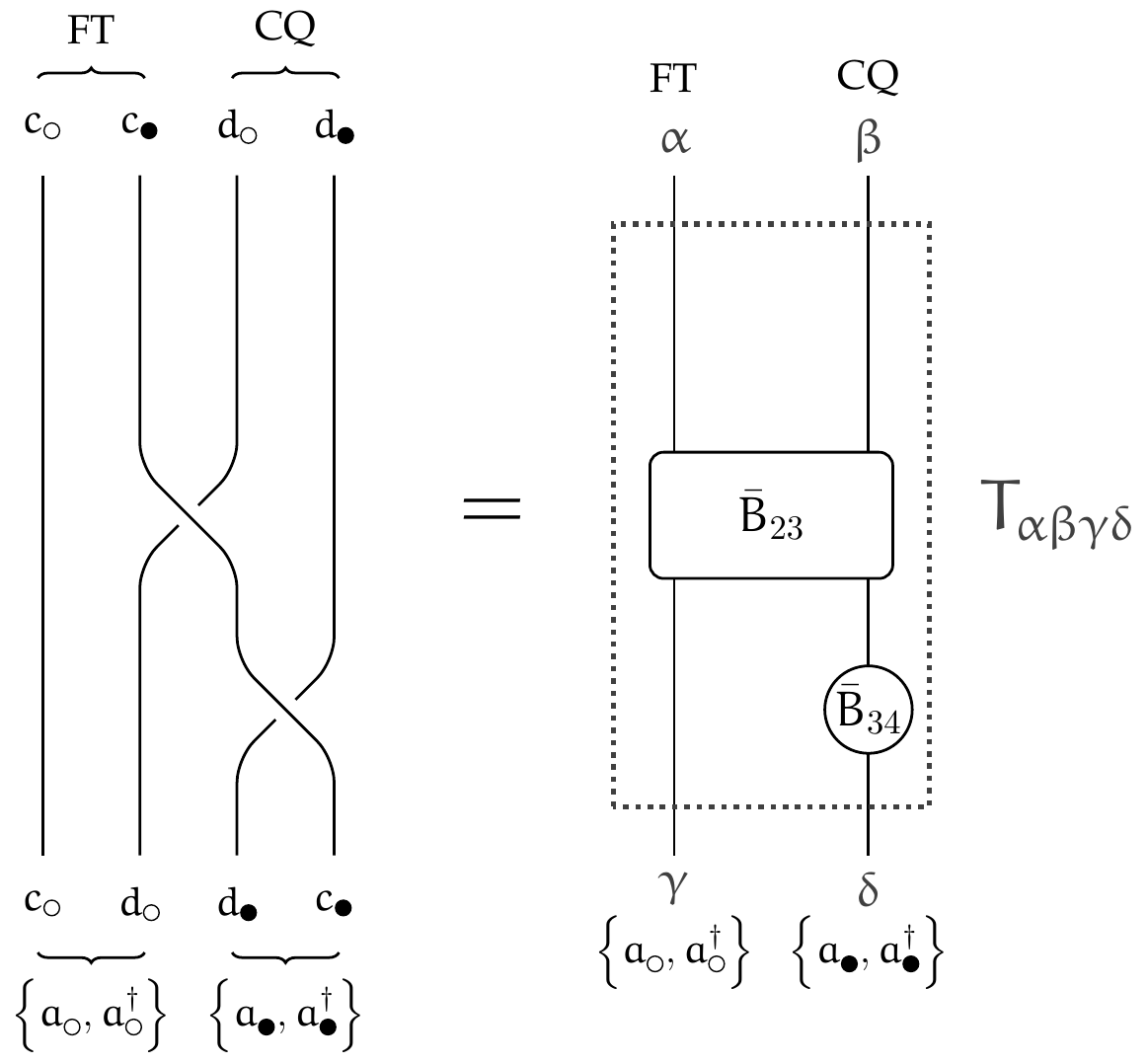}
	\caption{Majorana braiding operations to correctly recombine the different modes on the left, and the corresponding action on the Fock space of the two Dirac fermions on the right. FT refers to the modes out of the Fourier transformation, and CQ to the modes coming from the conserved quantities. The one-body gate braids Majorana modes within a Dirac fermion, whereas the two-body gate braids Majorana modes of different Dirac fermions. Valid patterns of clockwise/anticlockwise braidings are explained in the main text.}
	\label{fig:TikZ_Fun_MFwithinDF}
\end{figure}

The model has a large number of conserved quantities, which are accounted for by the occupation number of some ``vortex" Dirac fermions, and which now need to be reintroduced in the flow of the quantum circuit. Reintroducing these modes will lead, in fact, to having the correct number of original spins on the honeycomb lattice. The correct procedure is shown in Fig.(\ref{fig:TikZ_Fun_MFwithinDF}), where one Majorana mode of the Dirac fermion out of the Fourier transformation gets recombined with one Majorana mode of a ``vortex" fermion from the conserved quantities. Using the convention introduced before for the Majorana modes, we do the identification 
\beqa
\hat{c}_{\whitesite} = \hat{\gamma}_1 &~~~~& \hat{d}_{\blacksite} = \hat{\gamma}_3 \nonumber \\Ê
\hat{c}_{\blacksite} = \hat{\gamma}_2 &~~~~& \hat{d}_{\whitesite} = \hat{\gamma}_4 . 
\eeqa
Importantly, in order to reproduce the correct ground state of the Kitaev Honeycomb model, we observed that specific patterns of clockwise/antcilockwise braidings need to be used.

To be more specific, we have observed numerically that this \emph{depends on the conserved quantities}: if we choose the two-body ``external" braiding as anticlockwise, then the one-body gate in Fig.(\ref{fig:TikZ_Fun_MFwithinDF}) depends on whether the mode for the ``vortex" fermion is occupied (so that $\alpha_{\vec{r}} = -1$) or not (so that $\alpha_{\vec{r}} = +1$), according to 
\beqa
&\text{vortex mode} = \ket{0}, ~ \alpha_{\vec{r}} = +1 \Longrightarrow& \text{anticlockwise} \nonumber \\ 
&\text{vortex mode} ~= \ket{1},  \alpha_{\vec{r}} = -1 \Longrightarrow& \text{clockwise.} \nonumber
\eeqa
However, if we choose the two-body ``external" braiding as clockwise, then the one-body gate follows the opposite rule, i.e.,  
\beqa
&\text{vortex mode} = \ket{0}, ~ \alpha_{\vec{r}} = +1 \Longrightarrow& \text{clockwise} \nonumber \\ 
&\text{vortex mode} ~= \ket{1},  \alpha_{\vec{r}} = -1 \Longrightarrow& \text{anticlockwise.} \nonumber
\eeqa
These rules are important in order to reproduce the correct network for other eigenstates apart from the ground state. In the end, the overall procedure amounts for a two-body fermionic tensor which we call $T$, with the following matrix elements in all possible cases, with $\circlearrowright$ for clockwise and $\circlearrowleft$ for anticlockwise braidings:

\begin{enumerate}[(i)]

	\item ${\rm vortex ~ mode} = \ket{0}$, $B_{23} \circlearrowright$ , $B_{34} \circlearrowright$ 
	
		$T = \frac{1}{2} \begin{pmatrix} 1+i & 0 & 0 & -1-i \\ 0 & 1+i & 1+i & 0 \\ 0 & 1-i & 1-i & 0 \\ -1+i & 0 & 0 & 1-i \end{pmatrix}$
	
	\item ${\rm vortex ~ mode} = \ket{1}$, $B_{23} \circlearrowright$ , $B_{34} \circlearrowleft$ 
	
		$T = \frac{1}{2} \begin{pmatrix} 1-i & 0 & 0 & 1-i \\ 0 & 1-i & -1+i & 0 \\ 0 & -1-i & 1+i & 0 \\ 1+i & 0 & 0 & 1+i \end{pmatrix}$
	
	\item ${\rm vortex ~ mode} = \ket{0}$, $B_{23} \circlearrowleft$ , $B_{34} \circlearrowleft$ 
	
		$T = \frac{1}{2} \begin{pmatrix} 1-i & 0 & 0 & -1+i \\ 0 & 1-i & 1-i & 0 \\ 0 & 1+i & 1+i & 0 \\ -1-i & 0 & 0 & 1+i \end{pmatrix}$
	
	\item ${\rm vortex ~ mode} = \ket{1}$, $B_{23} \circlearrowleft$ , $B_{34} \circlearrowright$ 
	
		$T = \frac{1}{2} \begin{pmatrix} 1+i & 0 & 0 & 1+i \\ 0 & 1+i & -1-i & 0 \\ 0 & -1+i & 1-i & 0 \\ 1-i & 0 & 0 & 1-i \end{pmatrix}$
\end{enumerate}
 
\begin{figure}
	\includegraphics[width=.45\textwidth]{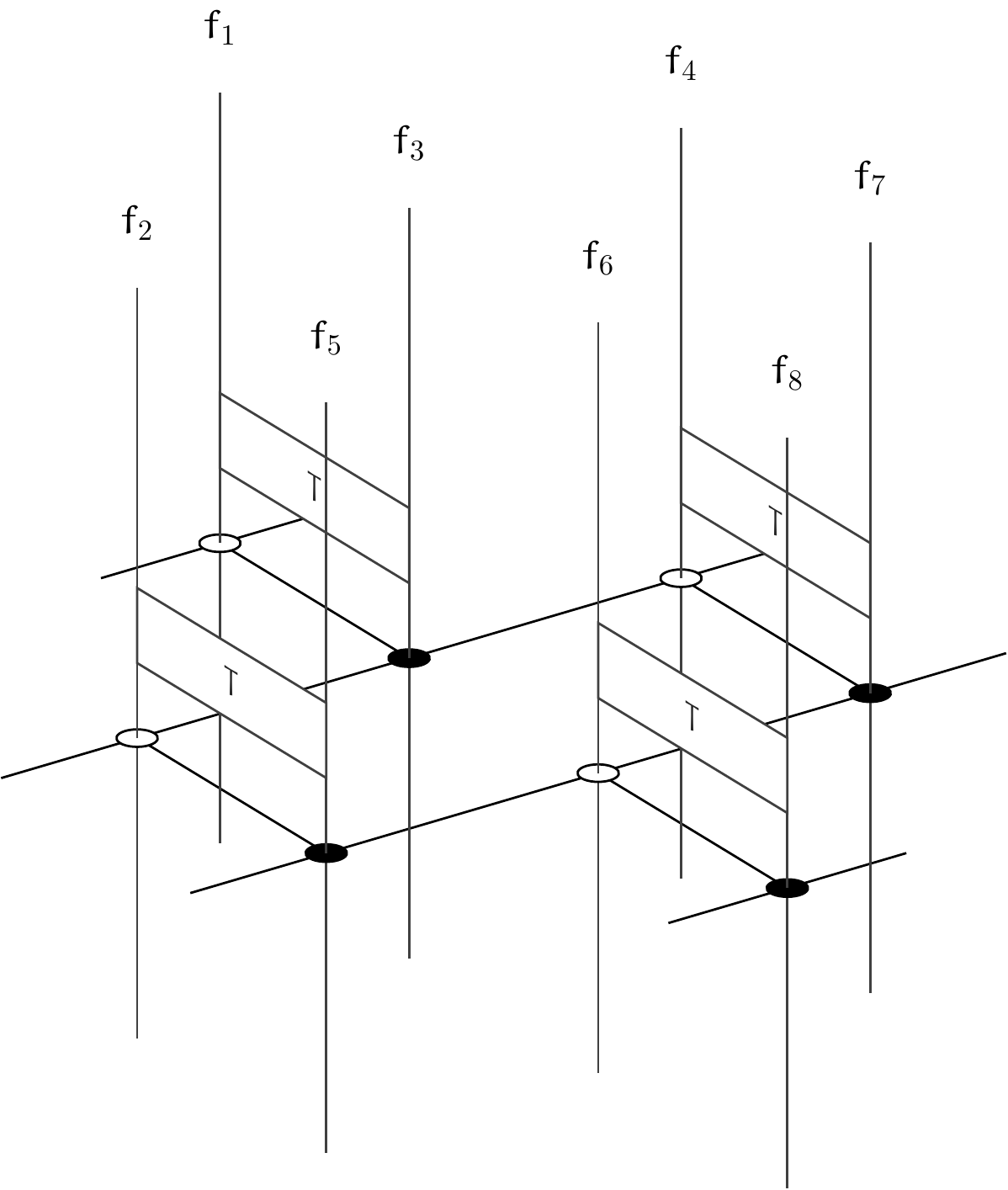}
	\caption{Fermionic tensor network to reintroduce the conserved quantities and braid the Majorana fermions. The transformation $T$ is given in Fig.(\ref{fig:TikZ_Fun_MFwithinDF}) for the case of an unoccupied ``vortex" fermion. Fermionic SWAPs are not included for simplicity.}
	\label{fig:TikZ_Fun_MajoranaNetwork_2}
\end{figure}

The fermionic tensors in Fig.(\ref{fig:TikZ_Fun_MFwithinDF}) act then locally on the $z$-links of the honeycomb lattice, as shown in Fig.(\ref{fig:TikZ_Fun_MajoranaNetwork_2}). Since this is a fermionic TN, all crossings of wires need to be accounted for by fermionic SWAP gates. In this diagram, modes $3$, $5$, $7$ and $8$ are the ``vortex" fermions. The other four fermions are the ones coming out of the Fourier transformation. Notice that, importantly, once the vortex fermion is fixed, the tensor $T$ becomes an isommetry, which is fundamental for some of the applications that we shall consider later.  

\begin{figure}
	\centering
	\includegraphics[width=.35\textwidth]{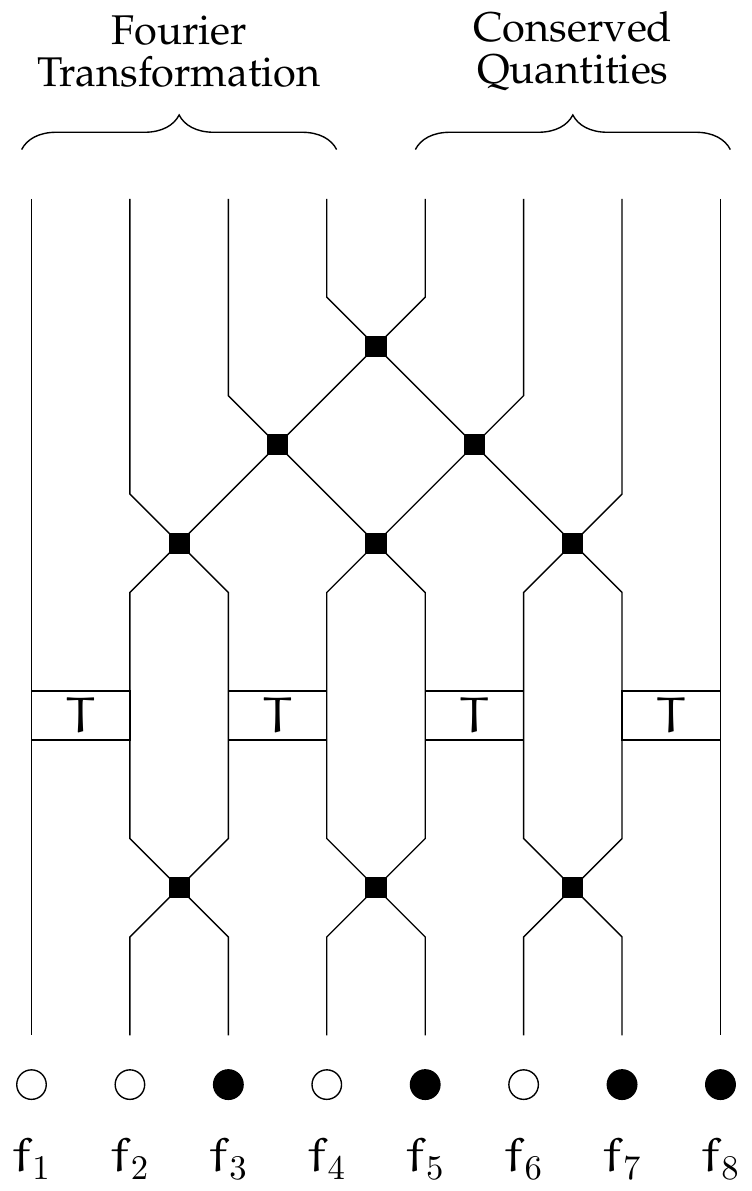}	
	\caption{A different projection of the TN in Fig.(\ref{fig:TikZ_Fun_MajoranaNetwork_2}), where wires do not cross any gate. The distinction between the modes coming from the Fourier transformation, and those coming from the conserved quantities (``vortex" modes), is also made explicit.} 
	\label{fig:TikZ_Fun_MajoranaNetwork_4}
\end{figure}

One needs to be careful in the implementation of the TN in Fig.(\ref{fig:TikZ_Fun_MajoranaNetwork_2}): to account for the correct fermionic SWAPS, it needs to be redrawn in such a way that no fermionic wire crosses a gate. To achieve this, one could, e.g., express the gate $T$ as a 2-site fermionic MPO with bond dimension 4, and the crossings would then be between wires only. Another option, however, is to simply project the network differently on the 2d plane of the paper. Whatever we do, the important thing is to ensure the correct fermionic ordering at the bottom of the network. As an example, the diagram in Fig.(\ref{fig:TikZ_Fun_MajoranaNetwork_2}) is redrawn in Fig.(\ref{fig:TikZ_Fun_MajoranaNetwork_4}), and represents the overall network needed to undo all the Majorana transformations in the Hamiltonian for an 8-site honeycomb lattice. It is also clear from the diagrams that this construction can be scaled up to lattices of any size. Again, we have also verified numerically that the overall TN up to here reproduces the correct eigenstates of the corresponding intermediate Hamiltonian. 

\subsection{Jordan-Wigner transformation} 
The final step is to bring back the fermions to spins on the lattice, which is implemented by undoing the Jordan-Wigner transformation. The surprising fact is that, somehow a bit counterintuitively, this transformation \emph{only changes the interpretation of the wires in the TN, and not the value of the overall coefficient obtained by the contraction}. This fact had already been noticed before \cite{JWIg}, and can be seen as follows: start from an arbitrary quantum state with an arbitrary number of spins-1/2 given by
\beq
\ket{\psi} = \sum_{n_1 n_2 \cdots } C_{n_1 n_2 \cdots } \ket{n_1 n_2 \cdots } \ , 
\eeq
with $n_i = 0,1$ for the down/up spin states at site $i$. This state can be rewritten as
\beq
\ket{\psi} = \sum_{n_1 n_2 \cdots } C_{n_1 n_2 \cdots } (\hat{\sigma}_1^+)^{n_1} (\hat{\sigma}_2^+)^{n_2} \cdots   \ket{\Omega} \ , 
\eeq
with $\ket{\Omega} = \ket{0 0 \cdots }$ (the state with all spins down. Applying the Jordan-Wigner transformation, one gets the fermionic state
\beq
\ket{\psi} = \sum_{n_1 n_2 \cdots } C_{n_1 n_2 \cdots } (\hat{a}_1^\dagger)^{n_1} (\hat{S}_1)^{n_2} (\hat{a}_2^\dagger)^{n_2} \cdots   \ket{\Omega} \ , 
\eeq
where $\hat{S}_i$ is the string operator attached to the creation operator $\hat{a}_{i + 1}^\dagger$, and is given by 
\beq
\hat{S}_i = \prod_{i' \leq i} (-1)^{\hat{n}_{i'}} = \prod_{i' \leq i} (1-2\hat{n}_{i'}) \ , 
\eeq
where the last equality can be easily checked by, e.g., Taylor-expanding $(-1)^{\hat{n}} = e^{i \pi \hat{n}}$. At this point, the state $\ket{\Omega}$ is reinterpreted as the vacuum of the fermionic modes. Moreover, it is also easy to check that (i) $\left\{ \hat{a}^\dagger, (-1)^{\hat{n}} \right\} =0$ and (ii) $\left[ \hat{a}_i^\dagger, (-1)^{\hat{n}_j} \right] = 0$ for  $i \neq j$. Therefore, using this we can write the state as 
\begin{widetext} 
\beq
\ket{\psi} = \sum_{n_1 n_2 \cdots } C_{n_1 n_2 \cdots } (-1)^{n_1 n_2} (-1)^{(n_1 + n_2)n_3} \cdots  (\hat{S}_1)^{n_2} (\hat{S}_2)^{n_3} \cdots  (\hat{a}_1^\dagger)^{n_1} (\hat{a}_2^\dagger)^{n_2} \cdots  \ket{\Omega} \,  
\eeq
\end{widetext} 
where the phases come from the commutation and anticommutation relations mentioned above. Finally, the action of the string operators is easily computed on the Fock state $(\hat{a}_1^\dagger)^{n_1}(\hat{a}_2^\dagger)^{n_2} \cdots  \ket{\Omega}$, and the result is the quantum state
\begin{widetext} 
\beqa
\ket{\psi}&=& \sum_{n_1 n_2 \cdots } C_{n_1 n_2 \cdots } (-1)^{n_1 n_2} (-1)^{(n_1 + n_2)n_3} \cdots (-1)^{n_1 n_2} (-1)^{(n_1 + n_2)n_3} \cdots   (\hat{a}_1^\dagger)^{n_1} (\hat{a}_2^\dagger)^{n_2} \cdots  \ket{\Omega}  \nonumber \\Ê
&=& \sum_{n_1 n_2 \cdots } C_{n_1 n_2 \cdots } (\hat{a}_1^\dagger)^{n_1} (\hat{a}_2^\dagger)^{n_2} \cdots  \ket{\Omega} \ , 
\eeqa
\end{widetext} 
where the last equation follows from the fact that the phases \emph{exactly cancel with each other}. The final state is nothing but a fermionic state in second quantization with \emph{the same}Ê coefficients $C_{n_1 n_2 \cdots }$ as the original state for spin-1/2. 

The result is then clear: the Jordan-Wigner transformation does not change the overall coefficient of the state, and therefore does not add anything quantitatively new to the TN. In our construction, this transformation is thus taken into account by \emph{simply saying that, from this point on, the wires in the TN represent spins instead of fermions}, so that we do not need to worry anymore about fermionic SWAP gates for the crossings. Mathematically, this is a site-by-site mapping from the Fock space of a spinless Dirac fermion to the Hilbert space of a spin-1/2. 

\subsection{Overall picture and summary}Ê

\begin{figure}
	\centering
	\includegraphics[width=.3\textwidth]{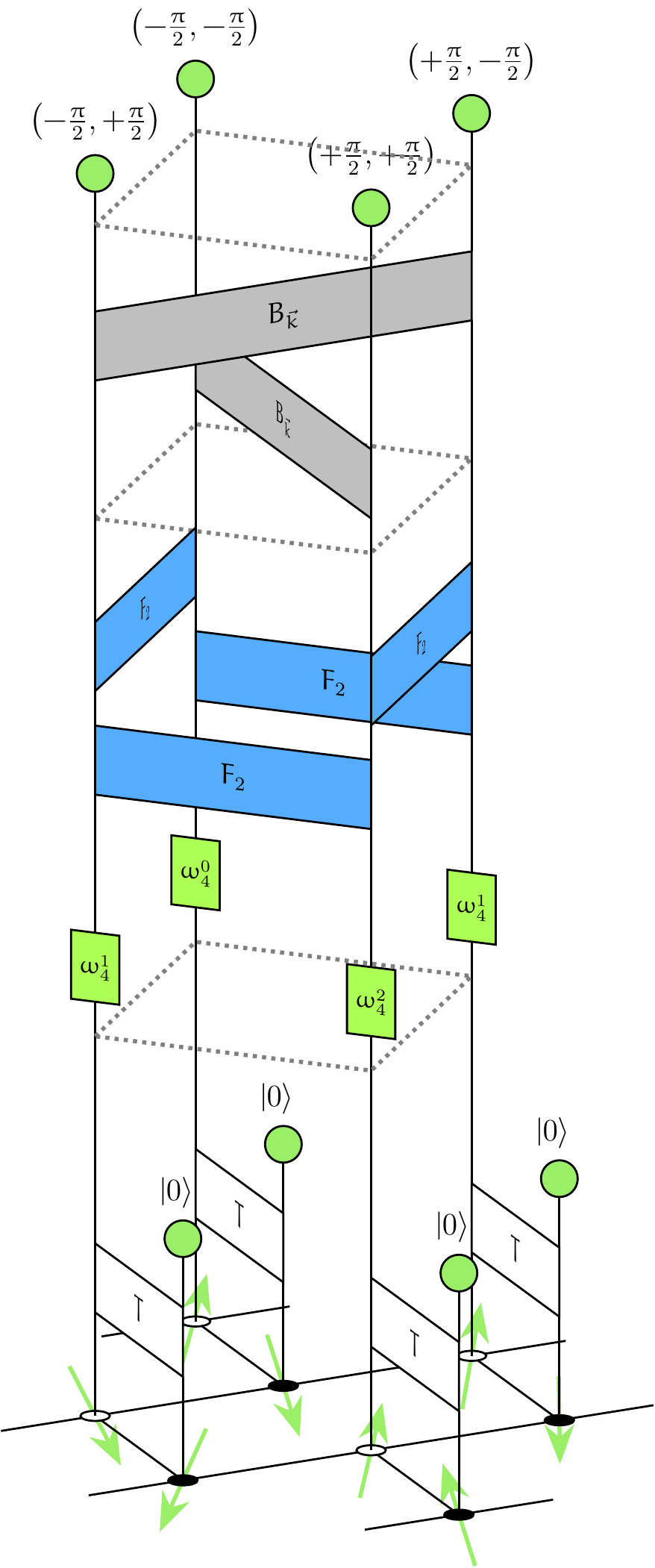}
	\caption{[Color online] 3d unitary TN for the ground state of the Kitaev model, for an 8-site honeycomb lattice with periodic boundary conditions. Physical degrees of freedom are spins, but essentially the whole network is fermionic. Fermionic SWAP gates are not included, for simplicity of the figure. Dotted lines are for reference.}
	\label{fig3d}
\end{figure}

\begin{figure}
	\centering
	\includegraphics[width=.25\textwidth]{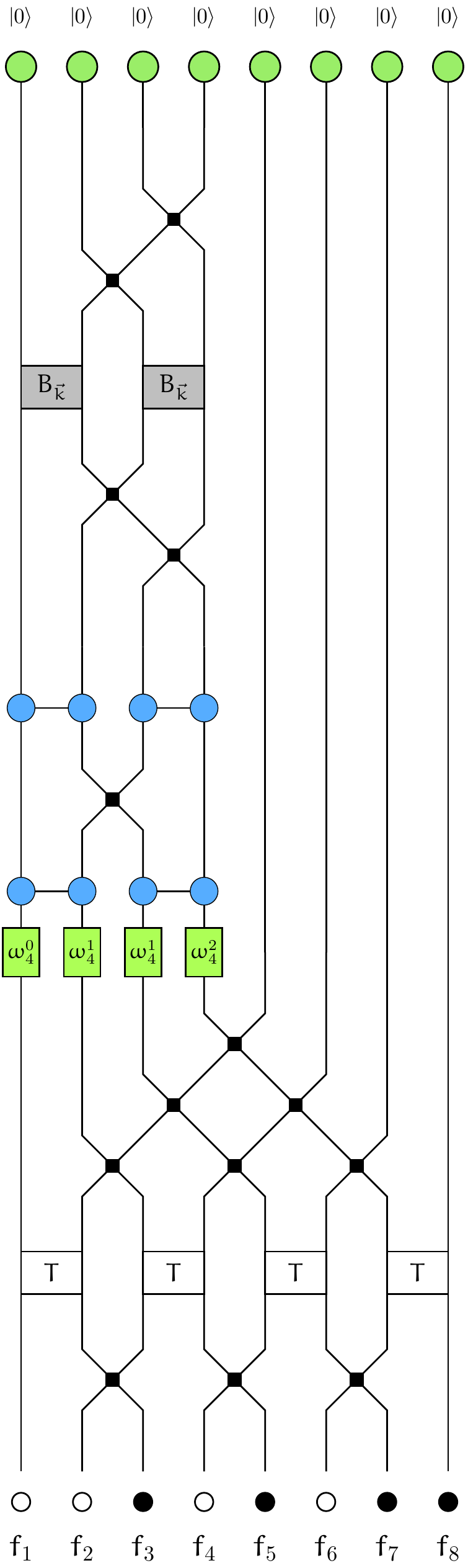}
	\caption{[Color online] Fermionic part of the TN in Fig.(\ref{fig3d}), but projected differently.}
	\label{fig2d}
\end{figure}

Following the steps described until now, the overall exact 3d unitary TN for the ground state of the Kitaev honeycomb model is represented in Fig.(\ref{fig3d}), for a small honeycomb lattice of 8 sites with periodic boundary conditions. The structure is easily scalable to the thermodynamic limit. In Fig.(\ref{fig2d}) we show an alternative ``2d projection" of the fermionic part of the same TN, more in the spirit of a quantum circuit. In what follows we review the basic steps leading to this construction. 

\subsubsection{Summary of the construction}

Let us now summarize the main ingredients leading to this exact 3d TN: 

\begin{enumerate}
\item{We start from a product state of Bogoliubov modes, and of ``empty" vortex modes (for the ground state).}
\item{The Bogoliubov transformation is implemented by 2-body unitary gates.}
\item{The Fourier transformation is implemented by a spectral TN.}
\item{Splittings and recombinations of Majoranas are implemented by a network of Majorana braidings.}
\item{Up to here, every crossing of indices (lines, wires) is fermionic, and is accounted for by a fermionic SWAP gate.}
\item{Finally, the Jordan-Wigner transformation simply maps each of the open fermionic indices to a spin (bosonic) index.}
\end{enumerate}

\subsection{Basic properties}Ê

\subsubsection{Arbitrary vortex sectors and arbitrary eigenstates} 

Arbitrary vortex sectors lead to a pattern of $\alpha_{\vec{r}}$ that is not necessarily translationally-invariant, which means that the resulting fermionic Hamiltonian cannot, in general, be diagonalized by a Fourier transformation. However, even if not invariant under translations, this fermionic Hamiltonian is still quadratic in the fermionic operators. It is well known that such Hamiltonians are always classically solvable in polynomial time, or in other words: formally, there is always a classical circuit with a polynomial number of gates that diagonalizes the Hamiltonian. Promoting such classical circuit to a (reversible, unitary) quantum circuit is always possible at a polynomial cost. Therefore, there is always a polynomial-size quantum circuit (say, of one- and two-body gates) that brings the Hamiltonian into diagonal form. Concerning our 3d TN construction, this quantum circuit should replace the Fourier and Bogoliubov transformations that were used to build up the TN for the ground state. Arbitrary eigenstates can therefore be easily constructed in this way, in a case-by-case basis depending on the vortex pattern. 

\subsubsection{Quantum circuit and causal cone} 

The unitary 3d TN that we just constructed is clearly an example of a polynomial-size quantum circuit (for a finite system) building up the quantum many-body state. This may have some interesting implications experimentally, e.g., to realize the ground state of a small Kitaev model in a small-size experimental setup with some quantum computing architecture.  But moreover, the fact that the network is made of unitary operators implies a \emph{causal cone structure} reminiscent to that in other TNs such as the MERA and the spectral TN \cite{MERA, specTN}. Its existence  means that expectation values of local operators only depend on the tensors inside the causal cone of the sites on which the operator is acting. This is a rather straightforward property of unitary networks, and is also the case in our construction. Notice, though, that unlike in the MERA, here the causal cone has no bounded width. Moreover, the fact that the 3d TN is constructed fully from unitary operators greatly simplifies the calculation of fidelities between states, as we show in the next section. 

\section{Applications}Ê
\label{sec4}

The 3d TN we just constructed allows for an easy understanding of several properties of the Kitaev honeycomb model, as well as a straightforward calculation of some relevant quantities. As an example, in this section we present three of them: the ground state fidelity diagram, the thermal fidelity in the vortex-free sector, and two-point correlation functions. 

\subsection{Ground state fidelity diagram}

In the thermodynamic limit, the ground state fidelity per site is a well-defined quantity that can be used to pinpoint different phases and phase transitions in quantum many-body systems \cite{fidelity}. For the Kitaev model, its derivative (the so-called ``fidelity susceptibility") was computed in Ref.\cite{fidKita}. Here we will show how this quantity is in fact straightforward within our formalism. 

\begin{figure}
	\centering
	\includegraphics[width=.4\textwidth]{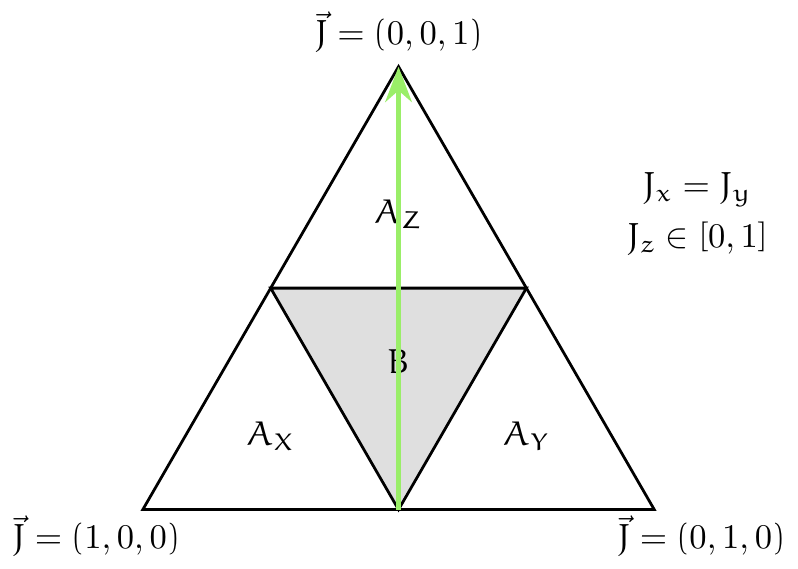}
	\caption{[Color online] Trajectory in the parameter space given by $J_x = J_y = 0.5 (1-J_z)$.}
	\label{figtranfid}
\end{figure}

Consider the Hamiltonian of the model for $J_x = J_y = 0.5 (1-J_z)$, so that $J_z$ is our single control parameter. The ground state fidelity $F$ is thus defined as 
\beq
F(\tilde{J}_z, J_z) =  |\langle \psi(\tilde{J}_z) | \psi(J_z) \rangle |^2
\label{fido}
\eeq
with $\ket{\psi(J_z)}$ the ground state for coupling $J_z$, and similarly for $\tilde{J}_z$. As shown in Fig.(\ref{figtranfid}), the phase transition between the A and B-phases is expected to take place at $(J_x, J_y, J_z) = (0.25, 0.25, 0.5)$. 

\begin{figure}
	\centering
	\includegraphics[width=.4\textwidth]{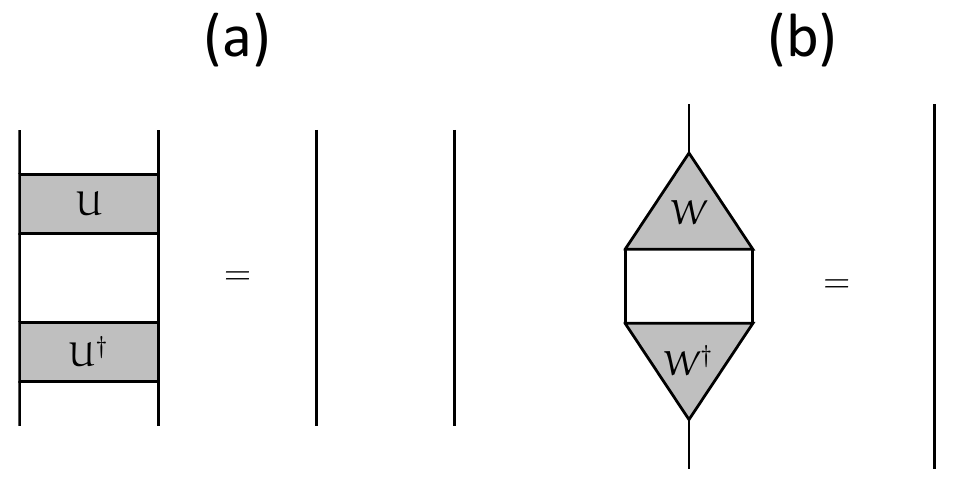}
	\caption{[Color online] Simplification in the contraction of (a) unitary, and (b) isometric tensors. Notice that the Majorana braiding operators $\hat{T}$ in our 3d TN are like in (b).}
	\label{cancel}
\end{figure}

\begin{figure}
	\centering
	\includegraphics[width=.18\textwidth]{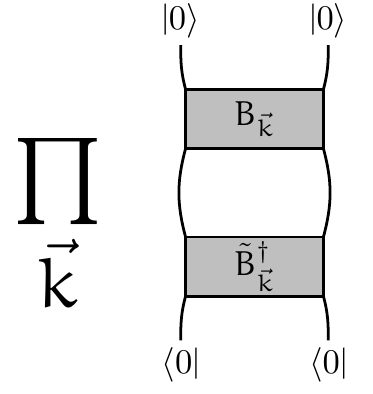}
	\caption{[Color online] TN diagram for the ground-state fideilty.}
	\label{figBogoFid}
\end{figure}

Using our 3d TN to compute the overlap in Eq.(\ref{fido}), it is easy to see that the couplings \emph{only enter in the Bogoliubov tensors} which are at the top of the structure, whereas all the other tensors are the same for both ground states. This has a nice implication: because of unitarity, \emph{the contraction of all the tensors cancels out (see Fig.\ref{cancel}) up to the Bogoliubov transformation}, leaving only the diagram in Fig.(\ref{figBogoFid}). In fact, all fermionic SWAPS also dissappear because it is a closed contraction. Thus, the expression for the ground state fidelity is given by 
\beq
F(\tilde{J}_z, J_z) = \prod_{\vec{k}} \left | \bra{0 0}Ê \tilde{B}^\dagger_{\vec{k}} B_{\vec{k}} \ket{00} \right|^2, 
\eeq
with $B_{\vec{k}}$ the Bogoliubov tensor for coupling $J_z$, $\tilde{B}_{\vec{k}}$ the one for $\tilde{J}_z$, $\vec{k}$ the momentum, and $\ket{00}$ the empty state of Bogoliubov fermions at momenta $\pm \vec{k}$. This expression corresponds to the TN diagram in Fig.(\ref{figBogoFid}). With a little bit of algebra one finds the expression 
\beq
F(\tilde{J}_z, J_z) =  \prod_{\vec{k}} \cos^2 (\theta_{\vec{k}} - \tilde{\theta}_{\vec{k}} ) , 
\eeq
with the Bogoliubov angles $\theta_{\vec{k}}$ defined in Sec.\ref{sec2}. This expression exactly matches the one in Ref.\cite{fidKita}, found by a different method. In fact, in the thermodynamic limit, we can find an exact analytical expression for the fidelity per site $d(\tilde{J}_z, J_z)$, which is defined as
\beq
\log d(\tilde{J}_z, J_z) \equiv \lim_{N \rightarrow \infty} -\frac{1}{N} \log F(\tilde{J}_z, J_z). 
\eeq
By replacing sums over momenta by integrals over the first Brillouin zone in the thermodynamic limit, it is not hard to see that  
\beq
\log d(\tilde{J}_z, J_z) = \frac{-1}{(2 \pi )^2} \int_{BZ} d^2\vec{k}\log \left (\cos^2 (\theta_{\vec{k}} - \tilde{\theta}_{\vec{k}}) \right). 
\eeq

\begin{figure}
	\centering
	\includegraphics[width=.5\textwidth]{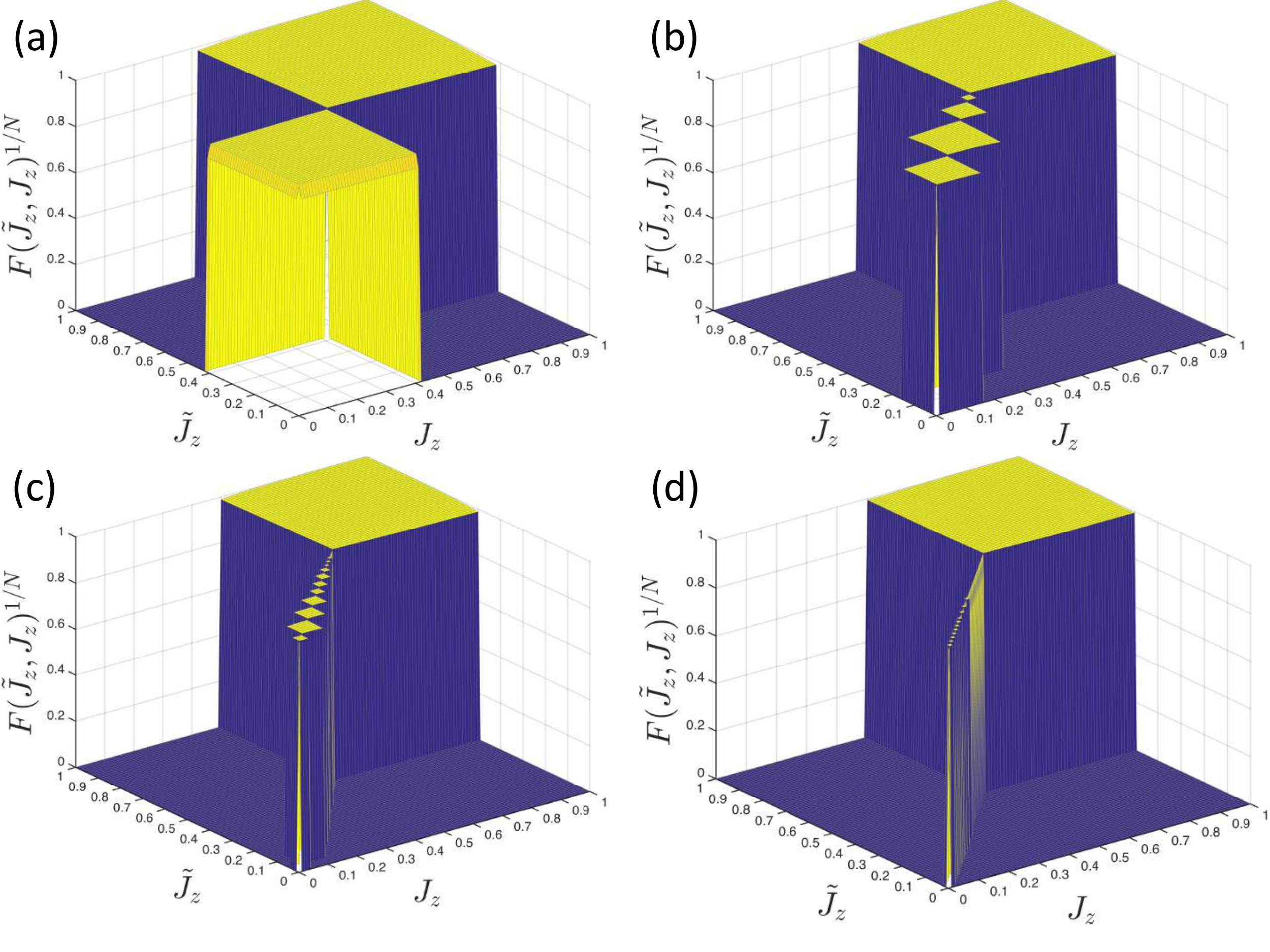}
	\caption{[Color online] Fidelity per lattice site for (a) 32, (b) 512, (c) 8192 and (d) 131072 spins on the honeycomb lattice, corresponding respectively to $4 \times 4$, $16 \times 16$, $64 \times 64$ and $256 \times 256$ square lattices of Bogoliubov momenta.}
	\label{FidData1}
\end{figure}

In Fig.(\ref{FidData1}) we show several calculations 32, 512, 8192 and 131072 spins on the honeycomb lattice ($4 \times 4$, $16 \times 16$, $64 \times 64$ and $256 \times 256$ square lattices of Bogoliubov momenta, respectively). One can there clearly see how finite-size effects tend to disappear when considering larger sizes, and how the diagram clearly pinpoints the transition between the A and B-phases, as expected. Unlike other continuous phase transitions, where one sees a pinch-point in the fidelity surface, here we see a sudden change in the fidelity and, in fact, the fidelity drops quickly to zero for any to different points in the B-phase. This fact may actually be related to the gapless nature of the B-phase and its infinite correlation length. 

\subsection{Thermal fidelity in the vortex-free sector}Ê

Here we show how, by using our 3d TN, it is also remarkably simple to compute the fidelity between thermal states restricted to the vortex-free sector \footnote{We focus on the vortex-free sector since our construction is explicitly valid. In other sectors, the resultant fermionic quadratic Hamiltonian may not be diagonalizable by a Fourier and Bogoliubov transforms, so that the TN construction may consist of a different fermionic circuit that should be computed on a case-by-case basis.}. Such states are given by
\beq
\hat{\rho}_\beta = \frac{\mathrm e^{- \beta \hat{H}_{\text{vf}}}}{Z_\beta}, 
\eeq
with $\beta$ the inverse temperature (in units where $k_B = 1$), $\hat{H}_{\text{vf}}$ is the Kitaev Hamiltonian restricted to the vortex-free sector, and $Z_\beta = \tr{Ê\left( \mathrm e^{- \beta \hat{H}_{\text{vf}}} \right) }$ the canonical partition function for this sector. The fidelity $F$ between two such operators is defined as 
\beq
F(\tilde{x}, x) = \tr{ \left( \sqrt{ \hat{\rho}_\beta^{1/2} \hat{\tilde{\rho}}_{\tilde{\beta}} \hat{\rho}_\beta^{1/2} } \right) } = \frac{\tr{ \left( \mathrm e^{- \beta \hat{H}_{\text{vf}}/2} \mathrm e^{- \tilde{\beta} \hat{\tilde{H}}_{\text{vf}}/2} \right) }}{\left( Z_\beta \tilde{Z}_{\tilde{\beta}} \right)^{1/2}}, 
\label{fidther}
\eeq 
with $x \equiv (J_x, J_y, J_z, \beta)$. 

\begin{figure}
	\centering
	\includegraphics[width=.4\textwidth]{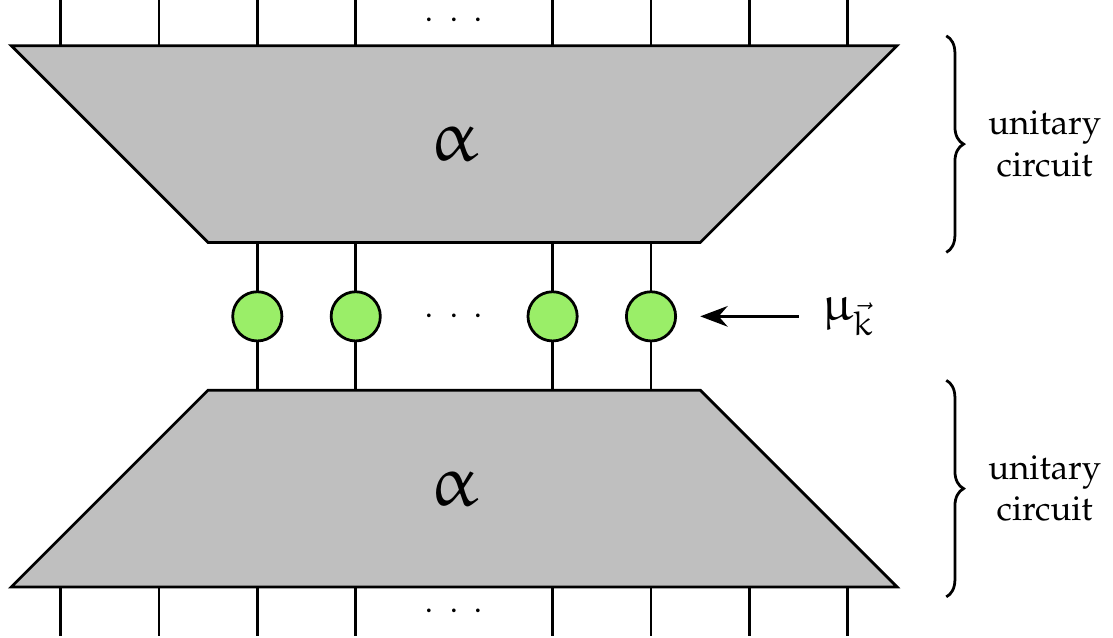}
	\caption{[Color online] TN diagram for $\mathrm e^{- \beta \hat{H}_{\text{vf}}/2}$, where state $\ket{\alpha}$ is our unitary 3d TN for the ground state, but for an excited configuration $\alpha$ of the Bogoliubov momenta.}
	\label{ThermalStateDiag}
\end{figure}

\begin{figure}
	\centering
	\includegraphics[width=.47\textwidth]{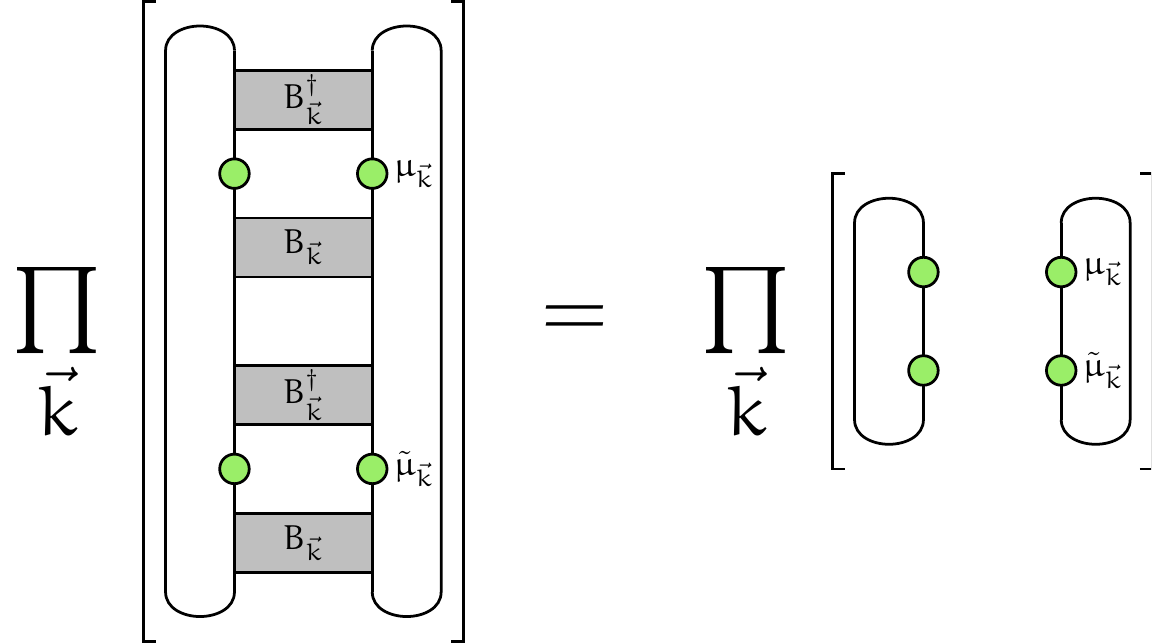}
	\caption{[Color online]. TN diagram for the numerator of Eq.(\ref{fidther}). For different couplings and temperatures, the left hand side is to be considered. If the couplings are the same and only temperatures differ, then this can be further simplified, and one obtains the diagram on the right hand side.}
	\label{ThermalFid2}
\end{figure}

In this sector, the exponential of the Hamiltonian admits a simple TN representation using the techniques, introduced before. To see this, first notice that 
\beq
\mathrm e^{- \beta \hat{H}_{\text{vf}}/2} = \sum_{\alpha \in \text{vf}} \mathrm e^{-\beta E_\alpha / 2} \ket{\alpha}Ê\bra{\alpha}, 
\eeq
with $E_\alpha$ the sum of the energies of the occupied Bogoliubov modes, and $\ket{\alpha}$ the corresponding eigenstate in this sector. In terms of the individual Bogoliubov energies $E_{\alpha, \vec{k}}$ at momentum $\vec{k}$, one has $E_\alpha = \sum_{\vec{k}} E_{\alpha, \vec{k}}$, and therefore 
\beq
\mathrm e^{- \beta \hat{H}_{\text{vf}}/2} = \sum_{\alpha \in \text{vf}} \prod_{\vec{k}}  \mathrm e^{-\beta E_{\alpha, \vec{k}} / 2} \ket{\alpha}Ê\bra{\alpha}.  
\eeq
This expression can be easily accounted for by the TN in Fig.(\ref{ThermalStateDiag}), where we introduced matrices $\mu_{\vec{k}}$ defined as 
\beq
\hat{\mu}_{\vec{k}} \equiv 
\begin{pmatrix}
			1 & 0	\\
			0 & \mathrm e^{-\beta E_{\alpha, \vec{k}}/2}
\end{pmatrix}
\eeq
in the $\ket{0}$ and $\ket{1}$ basis of Bogoliubov modes at momentum $\vec{k}$. It is thus easy to see that the only change in the TN between operators for $x \equiv (J_x, J_y, J_z, \beta)$ and $\tilde{x} \equiv (\tilde{J}_x, \tilde{J}_y, \tilde{J}_z, \beta)$ happens only in the Bogoliubov tensors, which include the couplings, and the $\hat{\mu}_{\vec{k}}$ tensors, which include the inverse temperature. This means that, again, when taking products of such operators, most of the tensors will cancel with each other because of the constraints in Fig.(\ref{cancel}). In the end it is straightforward to see that the TN diagram in Fig.(\ref{ThermalFid2}) corresponds to the numerator in Eq.(\ref{fidther}), whereas the denominator can be easily computed from 
\beq
Z_\beta = \prod_{\vec{k}} \tr \left( \hat{\mu}_{\vec{k}}^2 \right) \ . 
\eeq
For the case of equal couplings and different temperatures, the above derivations lead to the final expression
\beq
F(\tilde{\beta}, \beta) = \prod_{\vec{k}}Ê\frac{ \tr \left( \mu_{\vec{k}} \tilde{\mu}_{\vec{k}} \right) } {\left( \tr \left( \mu_{\vec{k}}^2 \right) \tr \left( \tilde{\mu}_{\vec{k}}^2 \right)\right)^{1/2}} \ , 
\eeq 
which in the thermodynamic limit leads to the fidelity per site
\beq
\log d(\tilde{\beta},\beta) = \frac{-1}{(2\pi)^2} \int_{BZ} d^2\vec{k}  \log \left( \frac{ \tr \left( \mu_{\vec{k}} \tilde{\mu}_{\vec{k}} \right) } {\left( \tr \left( \mu_{\vec{k}}^2 \right) \tr \left( \tilde{\mu}_{\vec{k}}^2 \right)\right)^{1/2}} \right). 
\eeq

\begin{figure}
	\centering
	\includegraphics[width=.5\textwidth]{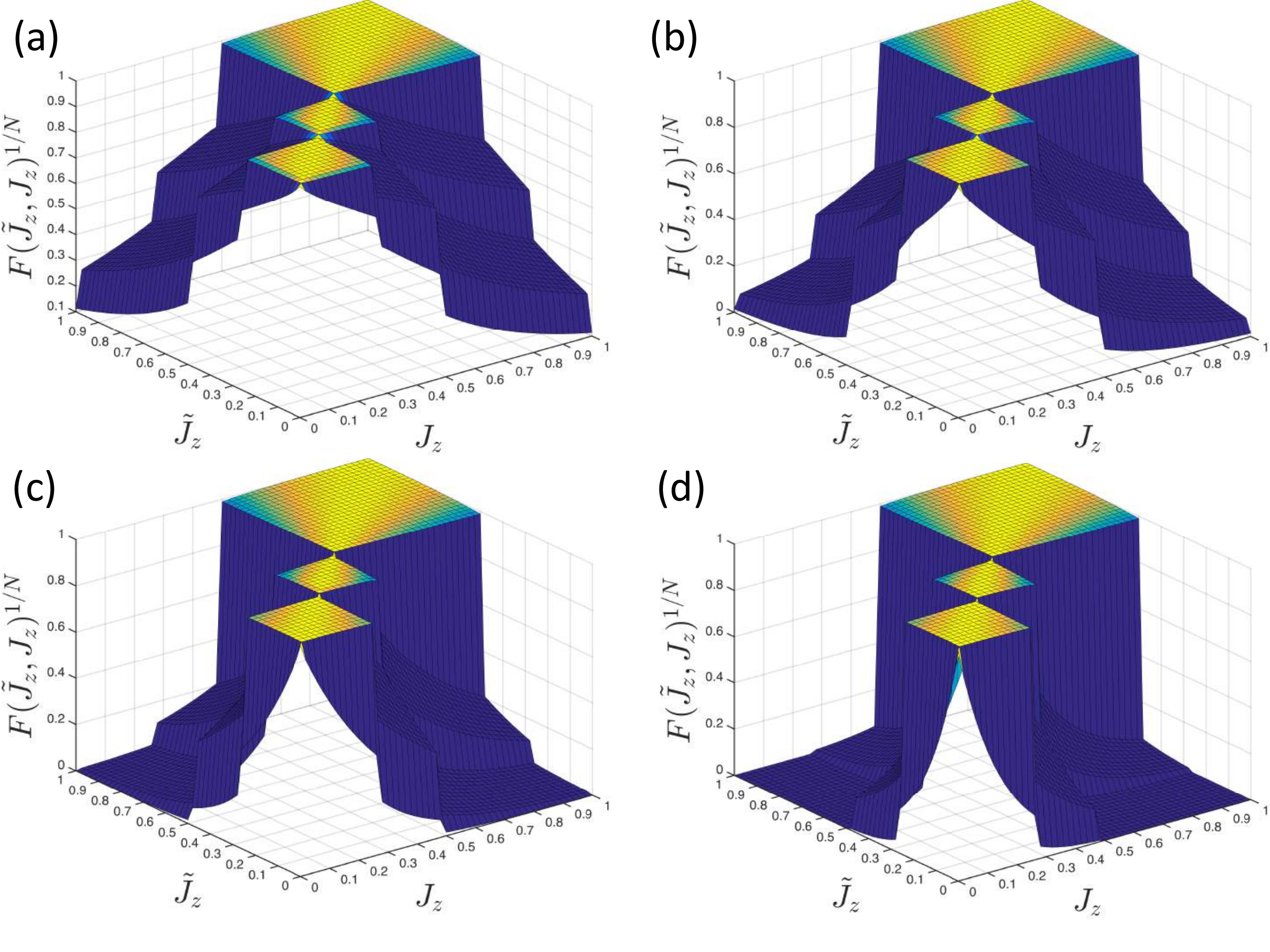}
	\caption{[Color online] Thermal fidelity per lattice site in the vortex-free sector for 128 spins ($8 \times 8$ square lattice of Bogoliubov momenta), along the trajectory in Fig.(\ref{figtranfid}), and inverse temperatures $\tilde{\beta} = \beta = $ (a) $50$, (b) $100$, (c) $200$ and (d) $400$.}
	\label{FidData2}
\end{figure}

\begin{figure}
	\centering
	\includegraphics[width=.5\textwidth]{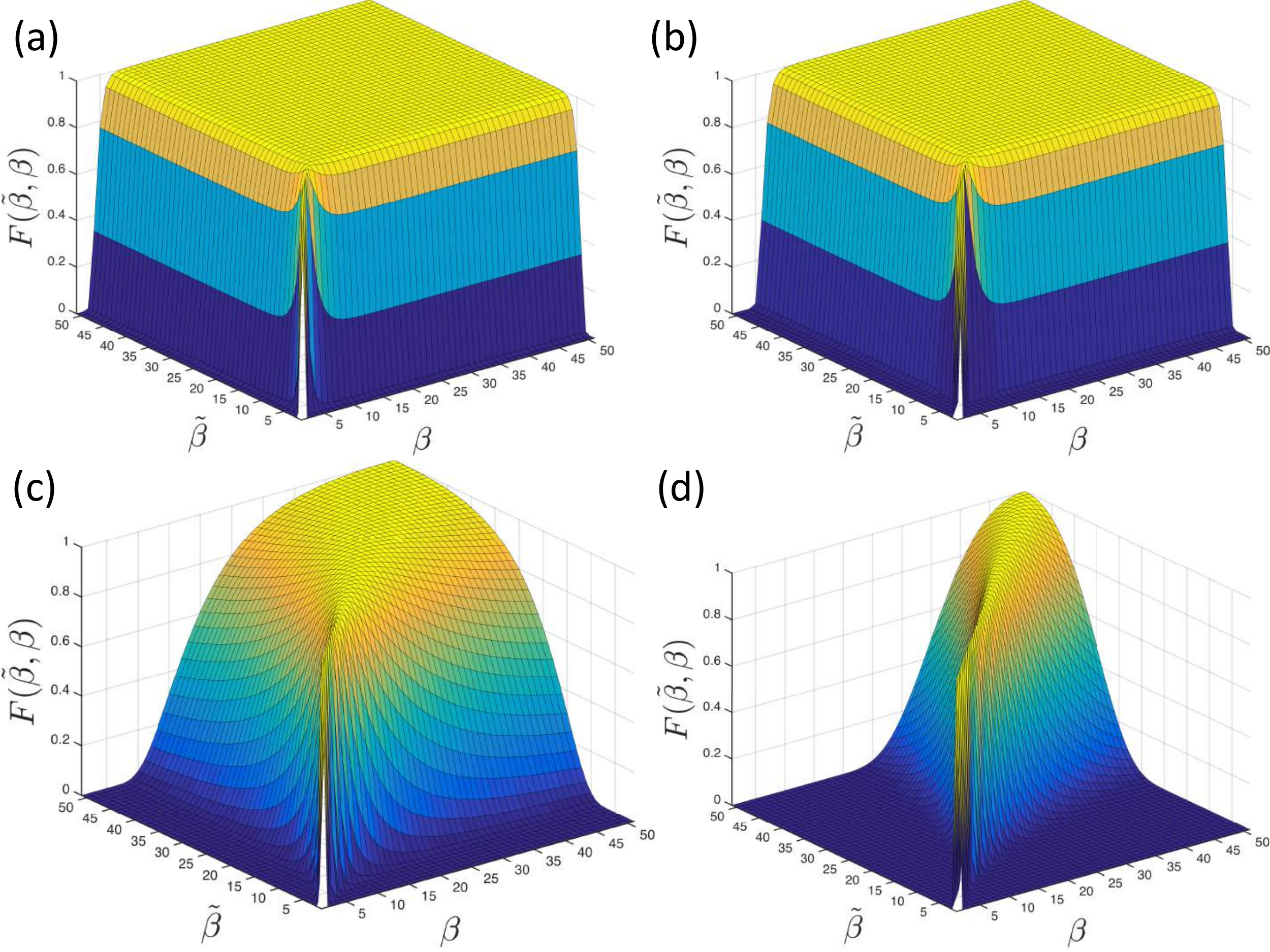}
	\caption{[Color online] Thermal fidelity in the vortex-free sector and different inverse temperatures for (a) 2048 spins ($32 \times 32$ Bogoliubov fermions) and $J_x = J_y = 0.1$, $J_z = 0.8$ (A-phase),  (b) 32768 spins ($128 \times 128$ Bogoliubov fermions) and $J_x = J_y = 0.1$, $J_z = 0.8$ (A-phase), (c) 2048 spins ($32 \times 32$ Bogoliubov fermions) and $J_x = J_y = J_z = 1/3$ (B-phase),  (d) 32768 spins ($128 \times 128$ Bogoliubov fermions) and $J_x = J_y = J_z = 1/3$ (B-phase).}
	\label{FidData3}
\end{figure}

Using these expressions we have computed the thermal fidelity in the vortex-free sector for different configurations of couplings, temperatures and sizes. In Fig.(\ref{FidData2}) we show the results per lattice site for 128 spins, couplings as in the trajectory of  Fig.(\ref{figtranfid}), and different fixed inverse temperatures. We observe a strong presence of finite-size effects also at non-zero temperature, with a clear different behaviour between ``thermal" A- and B-phases, which gets stronger as we approach lower temperatures. Complementary, in Fig.(\ref{FidData3}) we show the fidelity (this time not per lattice site) for fixed couplings and different inverse temperatures, for 2048 and 32768 spins, in the A- and B-phases. In the A-phase we see that finite-size effects are quite weak, whereas these become quite strong in the B-phase. 

As a final remark, let us stress that the thermal fidelity for other vortex sectors could also be computed in a case-by-case basis. As explained before, the 3d TN structure may change depending on the pattern of excited vortices, but is always unitary and of polynomial depth.  


\subsection{2-point correlation functions} 

Let us now consider the calculation of 2-point correlation functions. A well-known result for the Kitaev honeycomb model is that, in the static case (i.e., without time dependence), the spin-spin correlators of the ground state satisfy
\beq
\langle \sigma^{\alpha}_i \sigma^{\beta}_j \rangle = 
\begin{cases} 0 & \mbox{if } i,j\mbox{ not nearest neighbors} \\ g_\alpha \cdot \delta_{\alpha \beta} & \mbox{if } i,j\mbox{ nearest neighbors} \end{cases}
\label{corr}
\eeq
with $g_\alpha \neq 0$ only for an $\alpha$-type link, with $\alpha = x, y, z$. In other words: spin-spin correlators are only non-zero for nearest-neighbor sites, and between Pauli matrices of the same type as the interaction term for the considered link \cite{corrKita}. 

\begin{figure}
	\centering
	\includegraphics[width=.47\textwidth]{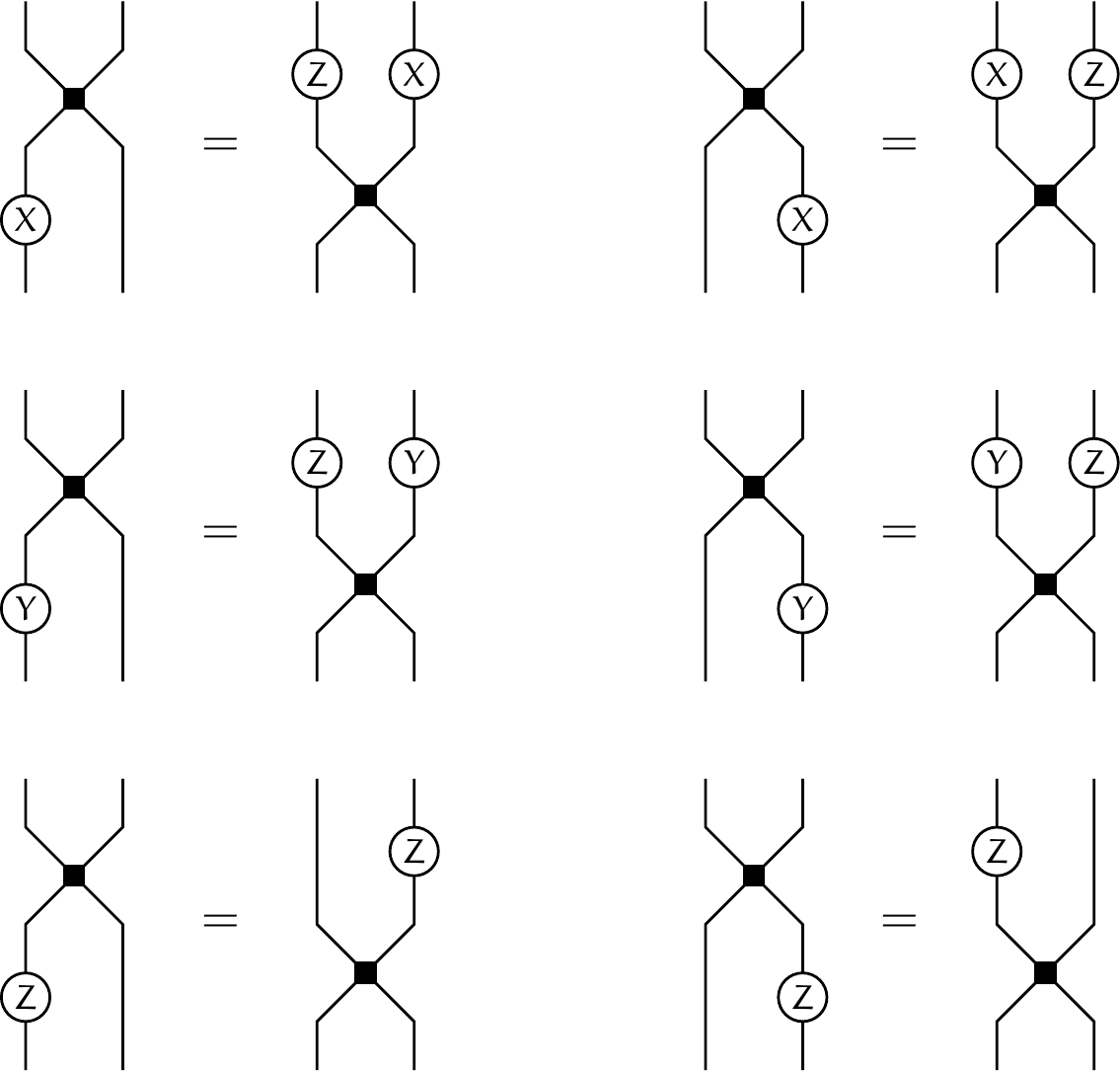}
	\caption{Several relations between the fermionic SWAP gate and operators $X$, $Y$ and $Z$. Notice that $Z$ is the fermionic parity operator.}
	\label{corr1}
\end{figure}

\begin{figure}
	\centering
	\includegraphics[width=.47\textwidth]{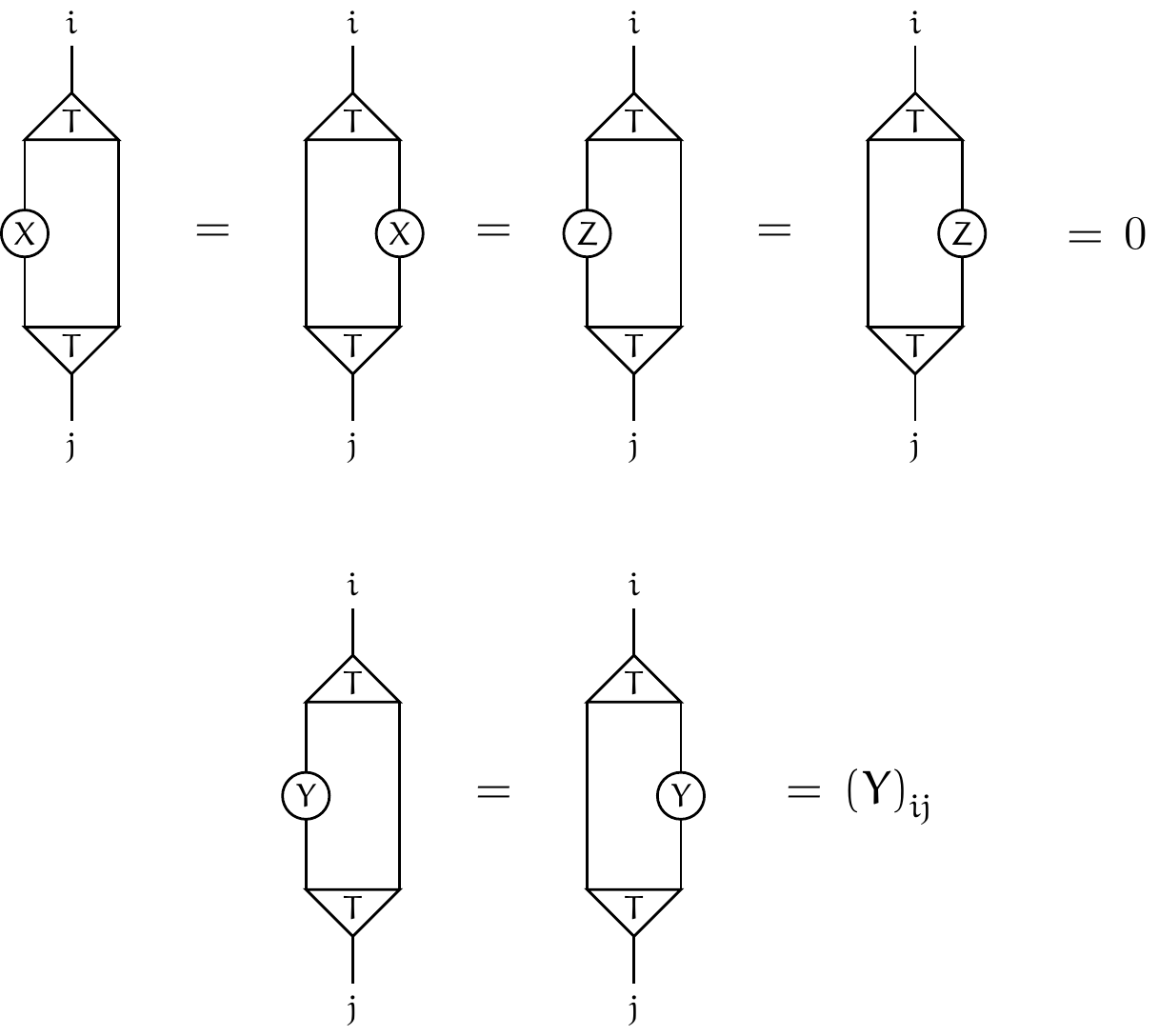}
	\caption{Matrix elements of operators $X$, $Y$ and $Z$ when sandwiched with the isometry $T$, resulting from the two-body gate $T$ after fixing one of its indices to $0$.}
	\label{corr2}
\end{figure}

\begin{figure*}
	\centering
	\includegraphics[width=1\textwidth]{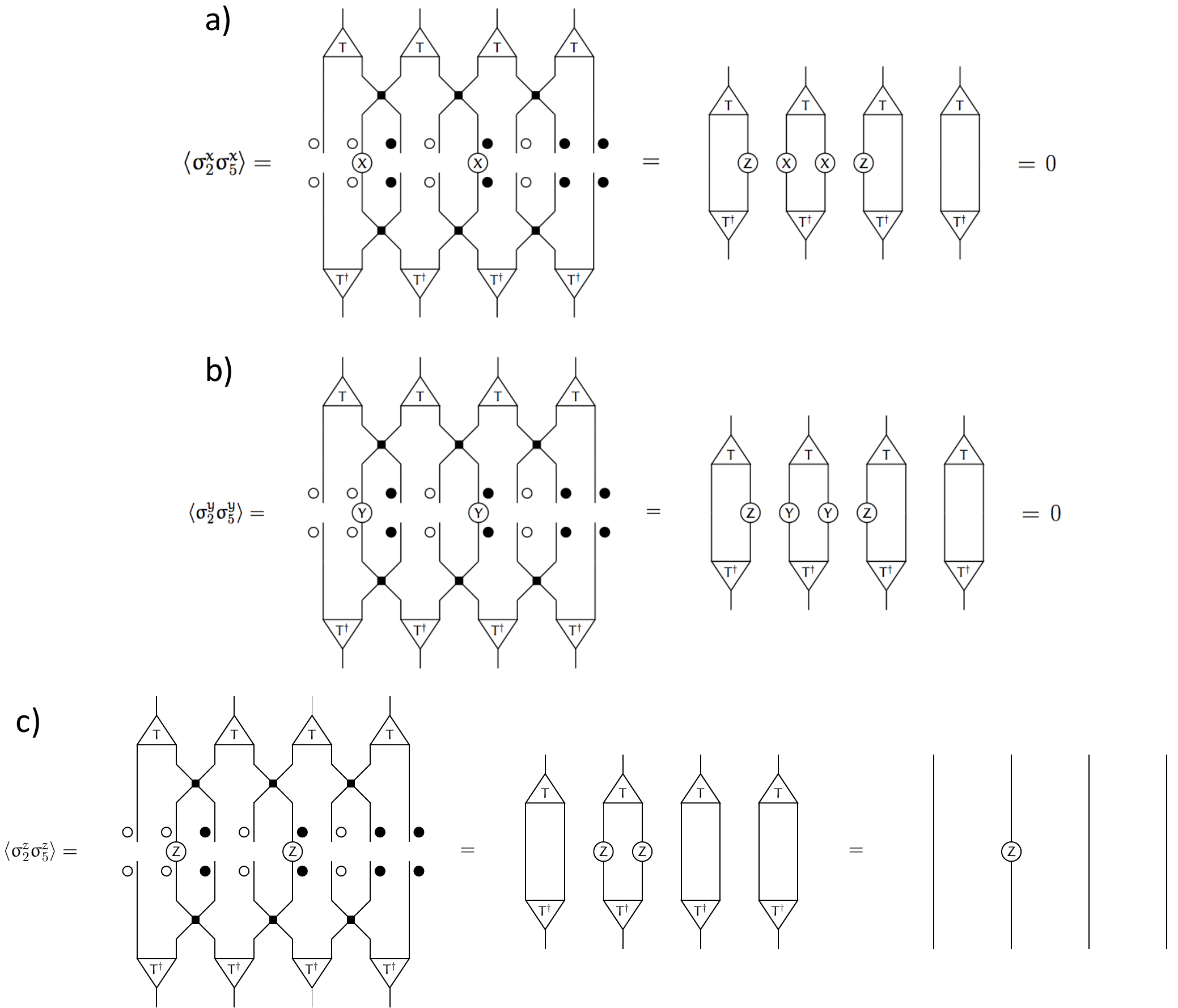}
	\caption{8-fermion example, in quantum circuit form (i.e., projected on the plane of the paper). For a $z$-link between fermions 2 and 5,  using the relations in Fig.(\ref{corr1}, \ref{corr2}), the only non-zero correlator is the one between Pauli-$z$ operators (case c).}
	\label{corr345}
\end{figure*}

With our TN construction this result is actually easy to reproduce. Because of the symmetry of the problem, we can focus on spin-spin correlation functions along the direction of the Jordan-Wigner string in Fig.(\ref{fig:JordanWignerPath}). Now, we can derive Eq.(\ref{corr}) by using the relations in the tensor network diagrams of Fig.(\ref{corr1}) and Fig.(\ref{corr2}), where we use the notation $X \equiv (\hat{a}^\dagger + \hat{a})$, $Y \equiv i (\hat{a}^\dagger - \hat{a})$, and $Z \equiv 2 \hat{a}^\dagger \hat{a} - 1 = (-1)^{\hat{n}}$. More specifically, it is easy to see that for sites that are not nearest neighbors there will always be a contribution coming from the diagrams in Fig.(\ref{corr2}) that  will multiply the expectation value by zero. For instance, correlators between Pauli-$x$ and $y$ matrices get a string of Pauli-$z$'s whose matrix elements when sandwiched with the isometry $T$ (i.e. the two-body gate $T$ for which one of the indices is fixed at $\ket{0}$) cancel exactly at every site along the string. For correlations between Pauli-$z$ operators there is no such a string, but the same is true at the distant sites for which the correlator is being computed. Next, focusing on nearest-neighbors, without loss of generality we can consider the spin-spin correlator for a $z$-type link since the rest follow by symmetry. The relations in Figs.(\ref{corr345}) imply directly that the only non-zero correlation function for such a $z$-type link can only be between Pauli-$z$ operators, and therefore Eq.(\ref{corr}) follows \footnote{In our formalism, though, it is not straightforard to extract the exact value of $g_\alpha$.}.  

\section{Implications for 2d PEPS} 
\label{sec5} 

We now would like to discuss the following question: is the 3d unitary TN presented before contractible, perhaps approximately, down to a 2d PEPS with finite bond dimension? As explained in the introduction, in the thermodynamic limit this should indeed be possible in the A-phase, as expected from the approximate mapping to the Toric Code Hamiltonian in this regime. In the B-phase, however, the situation is unclear: if such a description were  possible, then the PEPS constructed in this way would be PEPS for a gapless topological spin liquid (and with infinite correlation length) but fermionic bond indices, becoming chiral and gapped in the presence of a small magnetic field (or 3-spin) perturbation. This would be in part similar to some recently-found chiral fermionic PEPS \cite{Thorsten}, but with the ingredient of Dirac cones in the dispersion relation, and the first instance (as far as we know) of a gapped chiral topological 2d PEPS. However, if such a description were not possible, then this would be a practical example of an area-law state in 2d that cannot be represented by a 2d PEPS with finite bond dimension, in turn agreeing with the results in Ref.\cite{Jens}. Either way, answering this question would have important implications in the study of chiral topological order and TN states. 

In what follows we discuss the possibility of answering this question by using our 3d unitary TN, together with its pros and cons. This looks like a natural option but, as we shall see, is a non-trivial and hard task even with the aid of numerical TN methods. 

\subsection{Exact gaussian fermionic 2d PEPS as ground state is impossible} 

To begin with, we would like to review the fact that the ground state of the model cannot be written \emph{exactly} as a gaussian fermionic 2d PEPS \cite{fPEPS, Thorsten, comm}.  For this, a necessary condition is that the Hamiltonian after the Fourier transformation is given up to a constant by 
\beq
\hat{H} = \sum_{|\vec{k}| > 0} \left( \hat{d}^\dagger_{\vec{k}} ~ \hat{d}_{-\vec{k}} \right) \left( \sum_{i=1}^3 \sigma^i m_i(\vec{k}) \right) 	
		\begin{pmatrix}
			\hat{d}_{\vec{k}}	\\
			\hat{d}_{\text - \vec{k}}^\dagger
		\end{pmatrix}\ , 
\eeq
with $\sigma^i$ the $i$th Pauli matrix, and $m_i(\vec{k})$ trigonometric polynomials in $\vec{k}$ satisfying 
\beq
\left(m_1(\vec{k}) \right)^2 + \left(m_2(\vec{k}) \right)^2 + \left(m_3(\vec{k}) \right)^2 = \left(p(\vec{k}) \right)^2, 
\eeq
with $p(\vec{k})$ another trigonometric polynomial. Comparing this expression to Eq.(\ref{eq:BCSHamiltonian_2}), one can check easily that this is only true when two of the couplings in the Hamiltonian (e.g. $J_x$ and $J_y$) are zero. Therefore, the ground state of the model is, in general, not an \emph{exact} fermionic 2d PEPS with \emph{gaussian} projectors.

\subsection{Is an approximate fermionic 2d PEPS feasible?} 

In spite of the result from the above section, it may still be possible that the ground state of the model can be approximated with good accuracy by a (not necessarily gaussian) fermionic 2d PEPS. A possibility to check this would be, in fact, to approximate our 3d TN layer by layer by a 2d PEPS with finite bond dimension, and check if the final state approximates with good accuracy the exact ground state. This strategy is very tempting but, however, has the following important drawbacks:

\begin{enumerate}
\item{The Bogoliubov transformation is highly non-local in momentum space. In fact one can see that, when applied to the Bogoliubov vacuum, the resulting state typically obeys a volume-law for the entanglement entropy \cite{Noack}. This state is hard to approximate by a PEPS with finite bond dimension which, by definition, obeys an area-law. The subsequent fermionic Fourier transformation is also not guaranteed to produce a good approximation to the ground state on such a (drastically) approximated PEPS.}

\item{The fermionic Fourier transformation brings down the entanglement from the volume-law in the Bogoliubov state to an area-law which, in principle, could be handled by a PEPS. However, the state obtained from the exact contraction in our 3d TN would be a PEPS with a \emph{huge} bond dimension, and is numerically \emph{very} hard to approximate.}

\item{Additionally, the construction should be valid in the thermodynamic limit. But as the size gets larger, the approximations above become in fact harder since the Bogoliubov state becomes more  non-local. We are therefore restricted to small-size approximations.}  

\item{And finally, finite-size effects in the Kitaev honeycomb model are very strong, as has already been pointed out in the literature \cite{finiteKita}. Thus, the effect of a finite small-size may actually lead to a completely wrong conclusion about the thermodynamic limit.} 
\end{enumerate} 

The above does not mean that such a numerical approximation is impossible, but is definitely not obvious how to handle it with the above problems alltogether. As a matter of fact, we tried it for a system of 32 spins on the honeycomb lattice ($4 \times 4$ square lattice of Bogoliubov fermions), with inconclusive results. This finite size was pretty much close to our computer memory limit. Thus, the question remains open as to whether the ground state of the Kitaev honeycomb model is approximately a PEPS with finite bond dimension in the B-phase or not.   

\section{Conclusions and outlook}
\label{sec6}

In this paper we have derived a unitary TN with a 3d structure for the eigenstates of the Kitaev honeycomb model, focusing on the ground state. We have done this by expressing, in the TN language, every single step leading to the solution of this exactly-solvable model: Jordan-Wigner transformation, braiding of Majorana fermions, fermionic Fourier transformation, and Bogoliubov transformation. The produced TN allows for a straightforward calculation of several quantities of relevance, such as ground-state fidelity diagrams, fidelity between thermal states in the vortex-free sector, and properties of the two-point correlation functions. Moreover, we have discussed the possibility of approximating this 3d TN layer-by-layer in order to elucidate whether the ground state of the model in the B-phase can be approximated by a chiral 2d fermionic topological PEPS with finite bond dimension. Notice that several degrees of freedom in our construction could only be fixed by numerical checks for a finite-size system. This not only includes the signs in the Bogoliubov transformation, but also, e.g., the clockwise or anti-clokwise patterns of Majorana swaps. From the exact solution there was no apriori reason to choose one convention or the other, but this turns out to be relevant to reproduce the correct eigenstates in the TN construction.

The results in this paper can be extended to generalizations of the Kitaev honeycomb model to other lattices, spins and dimensions. The case of non-bipartite lattices would be particularly interesting, since time-reversal symmetry could be spontanously broken in such cases, thus leading to chiral ground states without the need of adding external magnetic fields \cite{KitaHon, nonbip}. Moreover, our results can also be a good starting point for numerical simulations of, e.g., Kitaev-Heisenberg \cite{KitaHeis} and Kitaev-Hubbard \cite{KitaHubbard} models with exotic TN structures. A similar 3d TN construction is also expected for the case of adding a 3-spin interaction to the model, which opens a gap and keeps the exact solvability. In fact, the property that 2d topological states show up at the surface of 3d TNs also occurs in other situations, e.g., in the so-called Walker-Wang models \cite{WW, hong}, so it would be interesting to see if there is any connection between these models and our results. Finally, the fact that our TN has the structure of a polynomial-size quantum circuit allows for direct realizations of these states of matter in quantum computing and quantum simulation architectures. 

\acknowledgements

We acknowledge fruitful discussions with Jens Eisert, Andrew J. Ferris, Didier Poilblanc, Matteo Rizzi, Hong-Hao Tu, Julien Vidal and Thorsten Wahl,  as well as funding from the JGU. 

\appendix 

\section{Construction of the finite-size Hamiltonian}
\label{appA}

\begin{figure}
    \includegraphics[width=.45\textwidth]{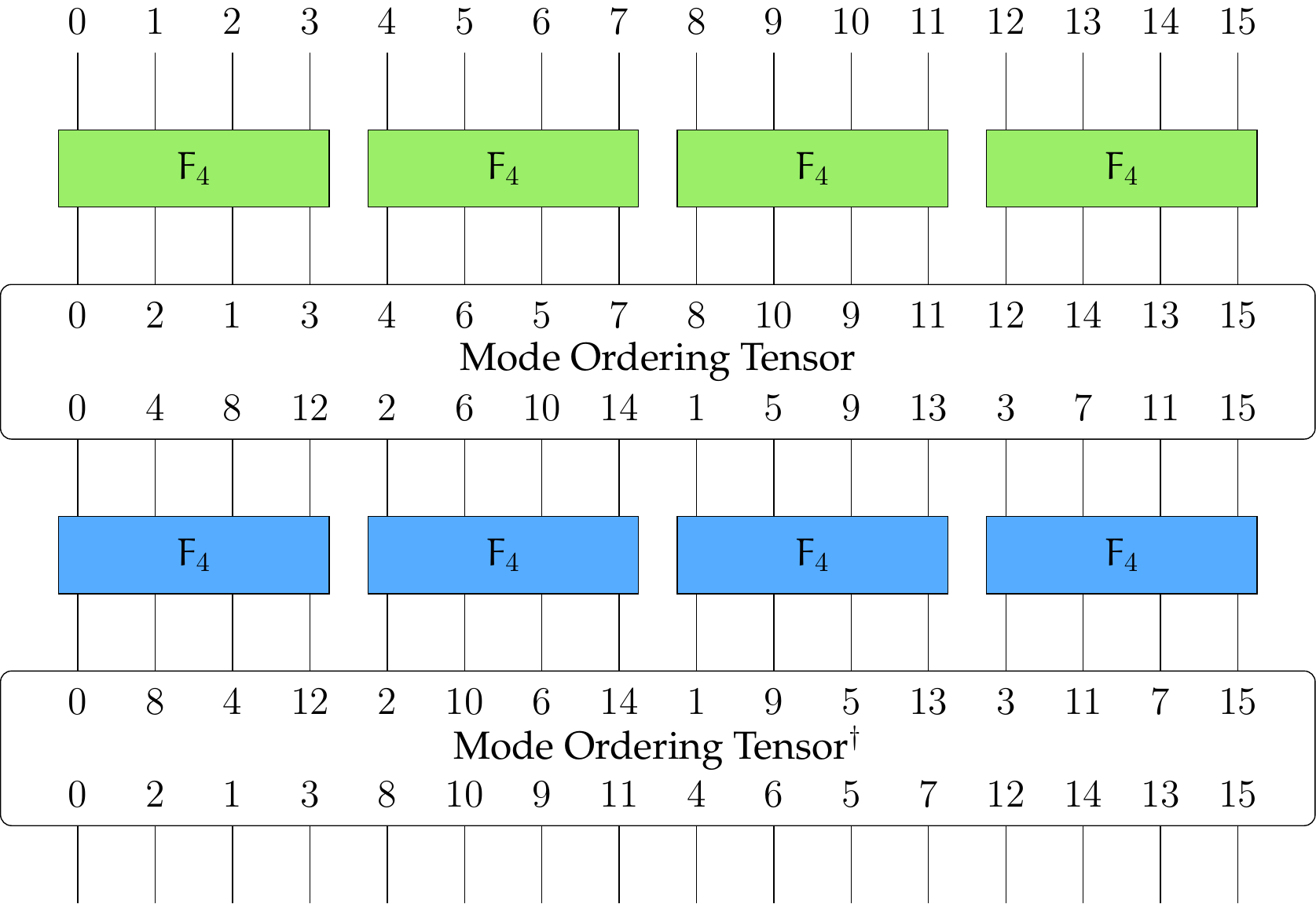}
  \caption{[Color online] Fourier transformation for a $4 \times 4$ square lattice, redrawn in a different way. Each crossing in the mode-ordering tensor as well as in the 4-site Fourier transformation $\hat{F}_4$ needs to be accounted for by fermionic SWAP gates.}
  \label{fig:TikZ_Fun_FourierNetwork_16Sites}
\end{figure}
\begin{figure}
	\includegraphics[width=.45\textwidth]{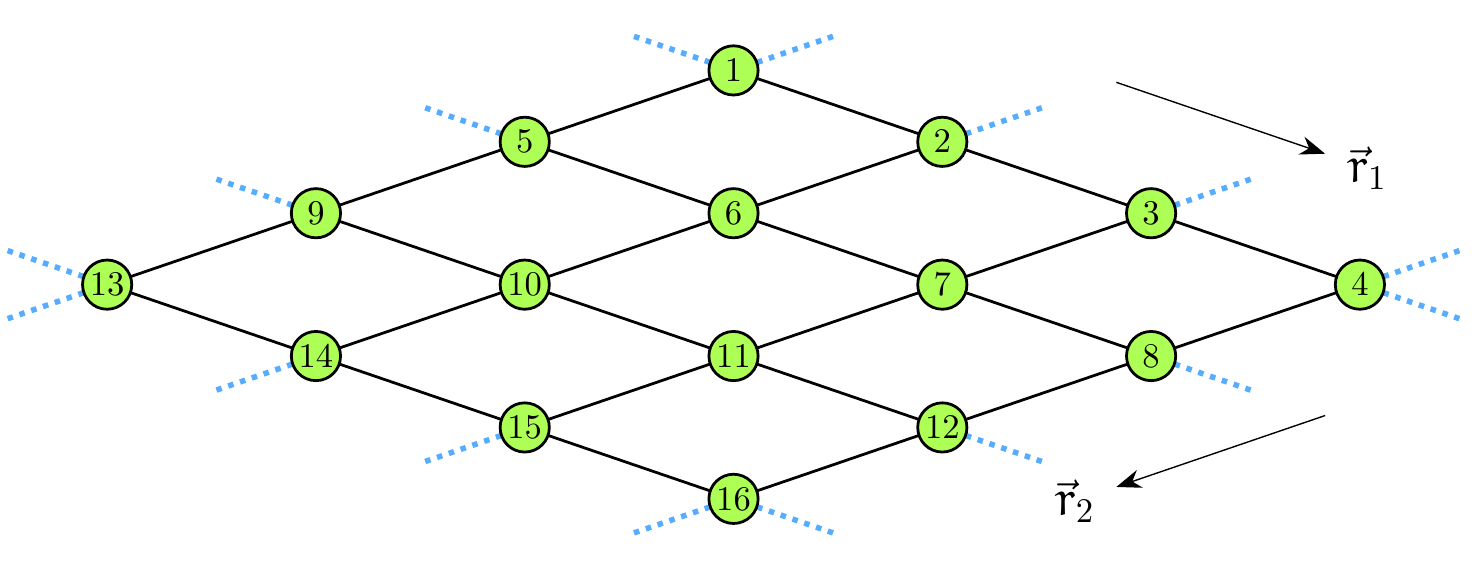}
	\caption{[Color online] $4 \times 4$ lattice in momentum space. The boundary conditions in both directions shape the lattice into a torus. Interacting terms at the boundary are in signalled in blue, and need to be treated differently in the Hamiltonian as compared to the bulk interaction terms, as explained in the text.}
	\label{fig:TikZ_Fun_MomentumSpace_Hamiltonian_Contribution_3}
\end{figure}

\subsection{After undoing the Fourier transformation} 

As explained in the main text, we have verified numerically that the TN state constructed up to this point is the exact eigenstate of the intermediate fermionic Hamiltonian in real space given in Eq.(\ref{fersq}), for a $4 \times 4$ lattice. There are however some subtilities. First, for easiness of the calculation it is worth redrawing the fermionic Fourier transformation as in Fig.(\ref{fig:TikZ_Fun_FourierNetwork_16Sites}), where the labelling of the modes is explicit. The 4-site Fourier transformation is denoted as $\hat F_4$ and already includes the bit-reversal operation. Crossings in this network are also accounted for by fermionic SWAP gates. The mode ordering tensor in the diagram sorts the modes to change the direction of the transformation, e.g., from $x$ to $y$ (this has to be undone once the Fourier tensors in $y$ direction have been applied). The overall network in this diagram performs the $4 \times 4$ Fourier transformation and sorts the modes as in Fig.(\ref{fig:TikZ_Fun_FourierNetwork_4x4Sites}). This procedure can be easily scaled up to arbitrary $N_x \times N_y$ sites, and therefore to the thermodynamic limit.  Second, we noticed that to do this exact verification, the finite-size Hamiltonian needs to be constructed carefully. This is because, due to the finite-size, the interaction terms at the boundary contribute with a negative sign in the Hamiltonian, unlike the bulk interaction terms. In practice this means that we need to work with antiperiodic boundary conditions, which is clear from the fact that
\beqa
	\hat{c}_{N_i+1} &=& \frac{1}{\sqrt{N_i}} \sum_{k_i} \mathrm e^{\frac{2\pi i k_i N_i}{N_i}} \mathrm e^{\frac{2\pi i k_i}{N_i}} \hat{c}_{k_i} \nonumber \\ 
	&=& - \frac{1}{\sqrt{N_i}} \sum_{k_i} \mathrm e^{\frac{2\pi i k_i}{N_i}} \hat{c}_{k_i} = - \hat{c}_{1}  
\eeqa
for both spatial directions and all possible values of the momentum. Fig. (\ref{fig:TikZ_Fun_MomentumSpace_Hamiltonian_Contribution_3}) 
shows the links that are affected in blue for the $4 \times 4$ square lattice. Taking this in to consideration it is possible to construct the appropriate finite-site version of the Hamiltonian in Eq.(\ref{fersq}), of which the intermediate TN constructed up to here is an exact eigenstate. When scaling up the system to the thermodynamic limit, this boundary effects become irrelevant, and the TN construction can also be extrapolated to this limit. 

\subsection{After undoing the Jordan-Wigner transformation}

The Jordan-Wigner transformation for a finite-size system leads to non-local interactions between degrees of freedom at the boundaries that need to be taken into account to verify the calculations. To correctly compensate for this effect in our numerical checks, we had to introduce boundary terms in the spin Hamiltonian with strings of $\hat{\sigma}^z$ attached, which cancelled out after the transformation, and reproduced the correct spectrum of the finite-size fermionic Hamiltonian.

\end{document}